\documentclass[twocolumn,times,tighten]{aastex631}
\usepackage{color}
\usepackage{enumitem}

\usepackage{hyperref}
\usepackage{verbatim}
\usepackage{amsmath,amstext,mathrsfs}
\usepackage[all]{hypcap} 
\usepackage{afterpage}    
\usepackage{subfigure}
\usepackage{xfrac}

\journalinfo{\hspace*{1pt}}
\defcitealias{LP00}{LP00}
\defcitealias{LP12}{LP12}
\defcitealias{GS95}{GS95}
\defcitealias{LV99}{LV99}
\defcitealias{KLP16}{KLP16}
\defcitealias{KLP17a}{KLP17}
\defcitealias{PaperI}{Paper I}
\DeclareFontFamily{OMS}{oasy}{\skewchar\font48 }
\DeclareFontShape{OMS}{oasy}{m}{n}{%
         <-5.5> oasy5     <5.5-6.5> oasy6
      <6.5-7.5> oasy7     <7.5-8.5> oasy8
      <8.5-9.5> oasy9     <9.5->  oasy10
      }{}
\DeclareFontShape{OMS}{oasy}{b}{n}{%
       <-6> oabsy5
      <6-8> oabsy7
      <8->  oabsy10
      }{}
\DeclareSymbolFont{oasy}{OMS}{oasy}{m}{n}
\SetSymbolFont{oasy}{bold}{OMS}{oasy}{b}{n}

\DeclareMathSymbol{\smallleftarrow}     {\mathrel}{oasy}{"20}
\DeclareMathSymbol{\smallrightarrow}    {\mathrel}{oasy}{"21}
\DeclareMathSymbol{\smallleftrightarrow}{\mathrel}{oasy}{"24}

\graphicspath{{./}{Fig/}}

\shorttitle{Statistics of polarization}
\shortauthors{Lazarian, Pogosyan \& Hu}
\submitjournal{ApJ}

\begin{document}

\title{Model of super-Alfv\'enic MHD turbulence and structure functions of polarization}

\author{A. Lazarian}
\affiliation{Department of Astronomy, University of Wisconsin-Madison, USA}

\author{D. Pogosyan}
\affiliation{Department of Physics, University of Alberta, Canada}

\author[0000-0002-8455-0805]{Yue Hu}
\affiliation{Institute for Advanced Study, 1 Einstein Drive, Princeton, NJ 08540, USA }

\begin{abstract}
MHD turbulence driven at velocities higher than the Alfv\'en velocity, i.e., super-Alfv\'enic turbulence, is widely spread in astrophysical environments, including galaxy clusters and molecular clouds. For statistical studies of such turbulence, we explore the utility of the polarization angle structure functions $D^\phi(R)= \left\langle\sin^2(\phi_1-\phi_2) \right\rangle$, where $\phi$ denotes the polarization angle measured at points separated by a projected distance $\mathbf{R}$ on the plane of the sky (POS). \citet{Laz_Pog22} showed that in the case of super-Alfv\'enic turbulence, the spectral slope of $D^\phi(\mathbf{R})$ differs from that of the underlying magnetic fluctuations, limiting its applicability for field strength estimation with known techniques. In this work, we provide an analytical framework that explains the modification of the $D^\phi(R)$ spectral slope in super-Alfv\'enic turbulence and validate our predictions with numerical simulations. We demonstrate that for super-Alfvénic turbulence, the structure function $D^\phi (R)$ gets shallower with the increase of $M_A$. Our study makes $D^\phi (R)$ a valuable diagnostic of super-Alfvénic turbulence and opens a way to obtain $M_A$ from observations. We also explore numerically the structure function of the polarization degree and the spectrum of the polarization directions, the latter being the Fourier transform of $D^\phi$. We discuss the implications of our findings for turbulence and magnetic field studies in the intracluster and interstellar media.
\end{abstract}


\section{Introduction}
\label{sec:intro}
Turbulent magnetic fields play a fundamental and multifaceted role in astrophysics (see \citealt{2004ARA&A..42..211E, MO07, 2019NatAs...3..776H, McKee_stone, 2022MNRAS.511..829H}). Observational evidence for magnetized turbulence comes from a wide variety of tracers, including interstellar density structures (e.g., \citealt{1995ApJ...443..209A, CL09, Xu_Zhang16}), velocity statistics \citep{1981MNRAS.194..809L, LP00, 2004ApJ...615L..45H, 2010ApJ...710..853C, chepurnov15, 2021ApJ...907L..40H, VDA, cattail, spectrum}, and synchrotron polarization \citep{LY18polar, ChepurZh}. Magnetized turbulence governs a range of high-energy and dynamical processes—it regulates the propagation, acceleration, and confinement of cosmic rays in galaxies and clusters \citep{YL02, Brun_Laz07, 2020PhRvX..10c1021M, 2025arXiv250507421H}, and influences every stage of star formation (e.g., \citealt{1956MNRAS.116..503M, 2006ApJ...647..374G, 2006ApJ...646.1043M, 2007IAUS..243...31J, 2022MNRAS.513.2100H, 2024MNRAS.532.2374M}).

The properties of magnetic fields varies because astrophysical turbulence (see \citealt{BeresnyakLazarian+2019}) spans a broad range of dynamical regimes. In galaxy clusters, it is typically super-Alfv\'enic \citep{Brun_Laz07, 2015MNRAS.450.4184Z, 2020ApJ...889L...1L}; in the solar wind, predominantly sub- to trans-Alfv\'enic \citep{solar_wind13, 2016ApJ...816...15W, 2020FrASS...7...83M, 2021ApJ...915L...8D, 2023arXiv230512507Z}; and in molecular clouds, trans- to super-Alfv\'enic \citep{padoan16_turb, 2019NatAs...3..776H, 2021ApJ...912....2H}. These regimes profoundly affect the structure, energetics, and evolution of astrophysical plasmas, shaping processes from star formation \citep{Mckee_Ostriker2007, McKee_stone} and cosmic-ray transport \citep{Jokipii1966, Jokipii_Parker1969, JokPar68, Sch94, YL02, Schlickeiser2002, Schilick12, LX21, Hucr22} to the morphology of magnetized media across scales \citep{Xu_Zhang_turb, Hu_mapping23}. In this paper, we focus on the properties of super-Alfv\'enic turbulence, a regime particularly relevant to the interstellar medium and galaxy clusters.

In this paper, we explore a turbulence measure based on the directional variation of the polarization angle. Previously, the dispersion of this measure—specifically, the dispersion of polarization angles (hereafter, PA)—was employed alongside the velocity dispersion in the Davis-Chandrasekhar-Fermi (DCF) technique (\citealt{Davis51, CF53}) to estimate the magnetic field strength. A more recent approach, the Differential Measure Approach (DMA; \citealt{Laz_Pog22}), introduces an alternative methodology for probing magnetic field strength. In this framework, the structure function of PA is defined as:
\begin{equation} 
D^\phi(\mathbf{R})=\frac{1}{2} \left\langle 1 - \cos(2(\phi_1-\phi_2)) \right\rangle=\langle \sin^2(\phi_1-\phi_2)\rangle,
\label{D} 
\end{equation}
where $\phi_1$ and $\phi_2$ are polarization angles measured at positions separated by a vector $\mathbf{R}$. This metric, when combined with the structure function of velocity centroids \cite{EL05}, allows localized estimates of magnetic field strength. A key advantage of Eq.~(\ref{D}) is its direct connection to Stokes parameters, making it readily applicable to observational polarization data (see also \citealt{2009ApJ...706.1504H}). For sub-Alfv\'enic turbulence, \citet{Laz_Pog22} established a theoretical connection between $D^\phi$ and the statistical properties of magnetic fluctuations. However, synthetic observations based on simulations of super-Alfv\'enic turbulence revealed a discrepancy: while the velocity statistics follow a Kolmogorov-like spectrum, the $D^\phi$ statistics exhibit a shallower spectral slope. The explanation of this inconsistency presents a theoretical challenge that we aim to address in this study.

Resolving this issue has significant implications for the interpretation of astrophysical turbulence and magnetic fields. Super-Alfv\'enic turbulence is believed to be common in galaxy clusters, where synchrotron polarization has been observed in diffuse radio structures, particularly in the outskirts of clusters \citep{Stuardi21,Hu_clusters24}. A deeper understanding of the behavior of $D^\phi$ across different turbulence regimes is therefore essential to extract reliable magnetic field information from polarization observations in such environments.

Studies of magnetized turbulence heavily rely on Faraday rotation measurements that require multi-frequency input. Polarization diagnostics of turbulence using spatial information (see \citealt{LP16}) were backed up by a mathematical description for sub-Alfv\'enic and trans-Alfv\'enic turbulence. A quantitative characterization of the $D^\phi$ measure in the super-Alfv\'enic regime expands the toolkit available to investigate turbulent magnetic fields. Our paper describes how the statistics of $D^{\phi}$ change with the Alfv\'en Mach number,
\begin{equation}
  M_A=V_L/V_A, 
  \label{eq:AMach}
\end{equation}
where $V_L$ is the turbulent velocity on the injection scale $L$ and $V_A$ is the Alfv'en velocity. Taking a broader outlook on the problem, we address the issues of obtaining information on the magnetic field in super-Alfv\'enic turbulence with dust and synchrotron polarization.

In what follows, in \S 2 we discuss the basics of super-Alfv\'enic turbulence, in \S 3  we describe our numerical simulations, in \S 4 we describe the statistics that we explore further in the paper, while in \S 5 we introduce the model describing the evolution of the polarized radiation in super-Alfv\'enic turbulence. The analytical expressions for the structure function of polarization angle are presented in \S 6. A comparison of the analytical and numerical results is provided in \S 7. In \S 8 we evaluate to what degree the statistics of polarized synchrotron and dust radiation reproduce the statistics of the projected magnetic field in super-Alfv\'enic turbulence and explore the synergy of the statistics of the polarization angle and the polarization degree. There we discuss the applicability of our results to addressing a challenging problem of obtaining magnetic field strength with observational data. The discussion of our results and the summary are provided in \S 9 and \S 10, respectively.

\section{Revisiting super-Alfv\'enic MHD turbulence}
\label{sec:basics}

To describe the statistics of $D^\phi$, a quantitative description of super-Alfv\'enic MHD turbulence is required. 
In what follows, we revisit the theory.

The theory of Kolmogorov hydrodynamic turbulence \citep{1941DoSSR..30..301K} is the most significant and fundamental advance of the turbulence theory. It can be understood using a mental picture of an eddy cascade with large eddies non-linearly evolving into smaller eddies, creating the cascade of turbulent energy from the injection to the dissipation scale. Kolmogorov turbulence is isotropic, as there are no preferred directions in an unmagnetized fluid. For a magnetized fluid, such a direction is the magnetic field direction. A deficiency of classical studies in \citep{1963AZh....40..742I,1965PhFl....8.1385K} was to consider magnetized turbulence as isotropic, nonetheless.
In addition, the MHD turbulence was assumed to be a pure wave-type phenomenon, similar to acoustic turbulence, which excluded strong nonlinearity, also an incorrect assumption. In total, the classical picture was shown to be untenable in the later research (\citealt{1981PhFl...24..825M}, \citetalias{GS95}). 

Within modern understanding of MHD turbulence (see \cite{BL19}), the MHD turbulence with magnetic perturbations $\delta B$ less than the mean $B$ is anisotropic and has a strongly non-linear regime. The turbulent cascade in compressible fluid can be approximated as a superposition of three cascades: Alfvénic, slow, and fast modes (\cite{CL03}).\footnote{We use the term "modes" rather than "waves" due to the strong non-linear interactions between perturbations. In particular, the Alfv\'enic modes decay on a timescale of the order of one period, which is not a periodic wave-like behavior.} 
The Alfv{\'e}nic turbulence cascade plays a dominant role in many processes. Alfv\'enic perturbations dominate the magnetic field wandering and variations of magnetic field directions (\cite{LV99}, henceforth LV99), which is the effect that we focus on in this study. The back-reaction of slow and fast modes on the Alfv{\'e}nic cascade is marginal in the strong Alfv{\'e}nic turbulence regime (\cite{GS95}, henceforth GS95, \citealt{2001ApJ...562..279L}, \citealt{CL02_PRL, CL03}) %see also appendix of \cite{2024arXiv240517985P}). 
Thus, for moderate sonic Mach numbers $M_s$, the scaling properties of Alfv{\'e}nic cascade change insignificantly with media compressibility \citep{CL03}. This fundamental property of the Alfv\'enic cascade enables us to apply the relations obtained for incompressible turbulence to astrophysically relevant settings.  

In this paper, we explore the turbulence with $\delta B$ on the injection scale larger than the mean $B$. The magnetic field at the injection scale is subdominant, but it becomes dynamically important at smaller scales where the turbulent velocities decrease. 

The Mach number $M_A$ given by Eq.~(\ref{eq:AMach}) is a useful measure to distinguish different regimes of turbulence.\footnote{Eq.~(\ref{eq:AMach}) provides the "velocity Alfv\'en Mach number", which is different from the "magnetic Alfv\'en Mach number" $M_{A,b}=\delta B/B$ for sub-Alfv\'enic turbulence, if turbulence is velocity driven \cite{Laz_Y_V25}. The difference between $M_A$ and $M_{A,b}$ disappears for super-Alfv\'enic turbulence.} 
MHD turbulence is sub-Alfv\'enic for the Alfv\'en Mach number $M_A<1$, trans-Alfv\'enic for $M_A\approx 1$, and super-Alfv\'enic for $M_A>1$. The original theory of the Alfv\'enic cascade of trans-Alfv\'enic turbulence was formulated in \citetalias{GS95} and was confirmed numerically in \cite{CV00, MG01, Bere14}. The formalism is also instrumental for describing the MHD cascade of super-Alfv\'enic turbulence. The super-Alfv\'enic turbulence can be roughly approximated as a sequence of two cascades, the hydrodynamic one at large scales, where the magnetic field plays a subdominant role, and the MHD one at smaller scales.\footnote{The theory of sub-Alfv\'enic turbulence \citep{LV99, Gal00} also has two regimes, but in both regimes, the magnetic field controls the turbulence dynamics. The properties of sub-Alfv\'enic turbulence depend on whether it is velocity or magnetically driven \citep{Laz_Y_V25}.}
The borderline of two cascades takes place when the turbulent motions injected as super-Alfv\'enic velocities at the large scale decrease in amplitude as a result of cascading and get equal to the Alfv\'en velocity at scale $l_A$ \citep{Lazarian06}. The corresponding scale of the transition can be obtained using the Kolmogorov scaling of hydrodynamic turbulence:
\begin{equation}
    v_l\approx V_L \left (\frac{l}{L}\right)^{1/3}.
    \label{v_GS}
\end{equation}
At scale 
\begin{equation}
    l_A\approx L M_A^{-3},
    \label{l_A}
\end{equation}
the turbulent velocity reduces to the Alfv\'en one, $V_A$. At this scale, the magnetic field gets dynamically important, and this scale can be viewed as the injection scale for trans-Alfv\'enic turbulence. Physically, this means that the injection scale $L^3$ contains $\sim M_A^{9}$ domains, each of which has a preferred magnetic field direction and exhibits an independent trans-Alf\'enic cascade. 

While the properties of the magnetic field within the $l_A$  domain are presented by the theory of trans-Alfv\'enic MHD turbulence (see Appendix A), the correlation of magnetic field directions in the $l_A$-domains depends on the initial conditions. If the magnetic field is generated by a turbulent dynamo (see \citep{XL16}) from the seed field with a correlation scale less than $l_A$, the directions of magnetic fields in different domains are not correlated. In numerical simulations of super-Alfv\'enic turbulence, the turbulence is initiated in the volume with the pre-existing large-scale magnetic field. In such settings, the residual correlation of the directions of the $l_A$-domains persists. This correlation decreases with the increase of $M_A$ and eventually vanishes for $M_A\rightarrow \infty$. This distinction related to the initial magnetic field structure has not been discussed earlier, and in this paper, we provide a quantitative model for the magnetic domain alignment.\footnote{We should point out that the turbulent dynamo is a part and parcel of super-Alfv\'enic turbulence. Turbulence, through the action of non-linear turbulent dynamo (see \cite{XL16}), tend to bring the magnetic and kinetic energies to equipartition at all scales. However, many astrophysical settings evolve on time-scales insufficient for the relatively slow non-linear turbulent dynamo to equalize the kinetic and magnetic energies.} 

In the presence of the initial mean magnetic field, its dynamical effect is counteracted by the dynamical pressure of turbulent motions perpendicular to the field.
The kinetic energy of the two magnetic field components is involved in this bending. The flow of fluid along the magnetic field also counteracts bending, inducing centrifugal force. To capture the underlying physics, one can approximate the distribution of magnetic field directions using a Boltzmann distribution, i.e. $\sim \kappa \sin^2 \theta/M_A^2$, where $\theta$ is an angle between the mean field and the magnetic field direction of a magnetic domain, $M_A^2$ represents the ratio of kinetic and magnetic energies, and $\kappa>1$ is a numerical factor reflecting the complex interaction of mean magnetic field and flow. In Appendix D, we provide a more sophisticated description of the alignment of magnetic domains in super-Alfv\'enic turbulence. This description is important beyond the immediate goal of the paper, i.e., the evaluation of the structure function $D^\phi (R)$. The transport processes in super-Alfv\'enic  i.e., transport of heat (\citep{Narayan_Medv, Lazarian06} and cosmic rays (\citep{Brun_Laz07, Brunetti17, Brunetti21}) require such a description. Those topics, however, are beyond the scope of our present study.

\section{Numerical simulation of super-Alfv\'enic turbulence in high-$\beta$ medium}

Our analytical predictions will be tested against results derived from 3D MHD numerical turbulence simulations. The numerical data sets were generated using AthenaK \citep{2024arXiv240916053S}, which solves the standard compressible ideal MHD equations under periodic boundary and isothermal conditions. The turbulence was driven solenoidally at a peak wavenumber of $2 \times 2\pi/L_{\rm box}$. The computational domain consists of a grid $792^3$, with numerical dissipation occurring on scales of approximately 10 cells. Details of the numerical setup, including the code and turbulence driving, are provided in \cite{2024MNRAS.527.3945H}. Table 1 summarizes the simulation parameters used to test our theoretical predictions.

\begin{table}
	\centering
	%\resizebox{\linewidth}{!}{
 \begin{tabular}{ | c | c | c | c | c |}
		\hline
		Run & $M_s$ & $M_A$ & $\beta$ & Resolution \\ \hline \hline
		A0 & 1.0 & 0.8 & 1.28 & $792^3$ \\ 
		A1 & 1.0 & 1.5 & 4.5 & $792^3$ \\
		A2 & 1.0 & 3.0 & 18.0 & $792^3$ \\ 
        A3 & 1.0 & 18 & 648 & $792^3$ \\
        \hline
  	%A0 & 1.21 & 1.25 & 0.51 - 1.56 & 0.58 - 1.53 & AthenaK\\\hline
	\end{tabular}
 %}
	\caption{\label{tab:sim} Setups of MHD turbulence simulations. The sonic and Alfv\'en Mach numbers, i.e., $M_s$ and $M_A$, are the instantaneous RMS values at each snapshot that is taken. $\beta = 2(M_A/M_s)^2$ is plasma compressibility.
 }
\end{table}

In our simulations, the sonic Mach number  $M_s \approx 1$. 
For super-Alfv\'enic turbulence this corresponds to the plasma $\beta = 2(M_A/M_s)^2>1$, 
The gas pressure is larger than the magnetic pressure.
This choice is justified by our interest in a magnetic structure that is not significantly modified by shocks, as we want to use the existing theory of MHD turbulence that does not include the effects of high $M_s$ shocks.
The case of turbulence in high-$\beta$ medium, similar to Kolmogorov turbulence, provides a valuable insight into more general cases. It is also directly applicable to particular astrophysical implications, i.e., to describing turbulence in galaxy clusters. 

\begin{figure*}[ht]
\label{fig:spectr_com}
\includegraphics[width=1.0\linewidth]{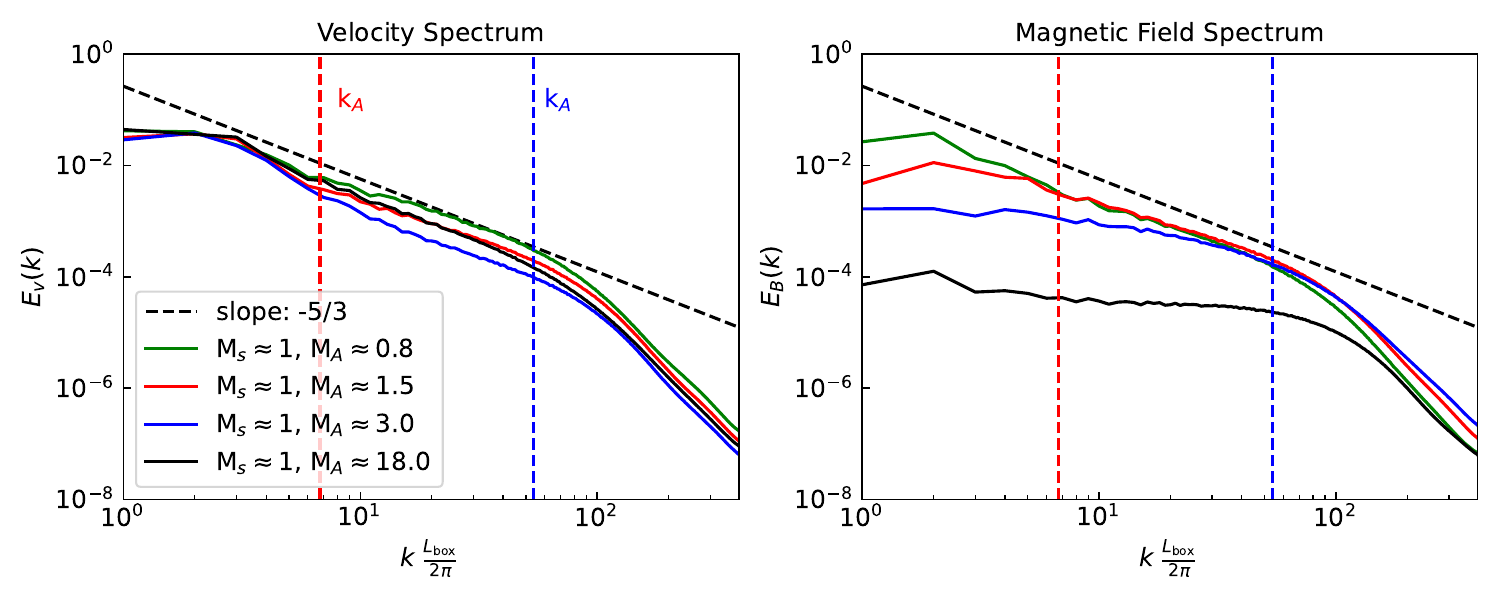}
\caption{Spectra of velocity (left) and magnetic field under different $M_A$ conditions. The dashed line represents the expected scaling for Kolmogorov turbulence $E_k \sim k^{-5/3}$ over the inertial range. The magnetic field spectrum in sub-Alfv\'enic regime is similar to that of velocity, but in super-Alfv\'enic regime, the spectrum becomes shallower, indicating that magnetic energy is subdominant for large scales. The dashed red and blue line represents the transition wavenumber $k_A$ for $M_A\approx1.5$ and $M_A\approx3.0$, respectively.}
\end{figure*}

The simulated magnetic and velocity spectra are shown in Fig.~\ref{fig:spectr_com}. Although the velocity spectra are Kolmogorov, $E_v(k) \propto k^{-5/3} $ for all $M_A$, for super-Alfv\'enic turbulence, the magnetic spectrum $E_B(k)$ is not. For scales smaller than $l_A$ (see Eq.~(\ref{l_A}), i.e., for $k>1/l_A$ it is Kolmogorov. For larger scales, i.e., for $k < 1/l_A$, $E_B(k)$ is flat as $M_A$ increases, tending to $E_B(k) \approx {\rm const}$. This corresponds to correlations on scales larger than $l_A$. These correlations are not described by a simplified theory of super-Alfv\'enic turbulence that we presented above. 

The spectrum of turbulence in the regime when magnetic field is dynamically unimportant is a contraversial issue. The Kazantzev theory of turbulent dynamo \cite{Kaza68} predicts the rising $\sim k^{3/2}$ spectrum of generated magnetic fields, provided that the initial seed magnetic field is small scale. In the simulations that we employ to test our theory, we use a weak uniform magnetic field. For $M_A=3$ we observe that the spectrum of magnetic field at scales when it is dynamically unimportant gets flat. We confirms this by performing the simulation for $M_A=18$. In this case the corresponding $l_A$ is much smaller than the dissipation scale $l_{diss}$ and the measured magnetic field spectrum is flat at all scales. Further in the paper, we use the assumption of the flat spectrum for weak large scale magnetic field of superAlfvenic turbulence.

\section{Structure Function of Polarization Angles}
\label{sec:Dphi_expand}

\subsection{Stokes parameters of polarized synchrotron and dust radiation}

Polarization measurements are traditionally used to study the Plane of Sky (POS) distribution of directions of astrophysical magnetic fields.\footnote{A recently introduced way of obtaining magnetic fields using Velocity Gradients \citep{LY18polar, 2018MNRAS.480.1333H,Hu_mapping23} and Synchrotron Intensity Gradients \citep{synch_grad17, Hu_clusters24} provides a powerful complementary tool.} Polarization is described by Stokes parameters $Q$ and $U$. For the synchrotron radiation we have
\begin{equation}
\begin{aligned}
Q &\propto  \int\! dz \; n_{e} | B_\perp |^{\alpha} \; (B_x^2 - B_y^2) \\
U & \propto  \int\! dz \; n_{e} | B_\perp |^{\alpha} \; 2 B_x B_y
\label{eq:stokes_dust}
\end{aligned}
\end{equation}
where $\alpha$ is the synchrotron index determined by the relativistic electron power spectrum, $B_x,B_y$ are the magnetic field components in $x$ and $y$ directions, and $z$ is along the line of sight. 

The study of synchrotron statistics in \citealt{LP12}) shows a marginal sensitivity of the result to the index $\alpha$
and that to study correlation scalings $\alpha=0$ can be adopted.
We can look at Eq.~(\ref{eq:stokes_dust}) in polar coordinates for the magnetic field, $B_\perp=B \sin\gamma$, 
$B_x = B \sin\gamma\cos\varphi, B_y = B \sin\gamma \sin\varphi, B_z = B \cos\gamma$. For $\alpha=0$
\begin{equation}
 \begin{aligned}
Q &\propto  \int\! dz \; n_{e} B^2 \;  \sin^2\gamma(z) \cos 2\varphi(z)  \\
U & \propto  \int\! dz \; n_{e} B^2 \;  \sin^2\gamma(z) \sin 2\varphi(z)
\label{eq:stokes_sync_angles}
\end{aligned}
\end{equation}
to recognize that we deal with the LOS averaged orientation of the magnetic field.

Dust emission can also be described using formulas similar to Eq.~(\ref{eq:stokes_sync_angles})
\begin{equation}
 \begin{aligned}
Q &\propto  \int\! dz \; n_{\rm dust}  \sin^2\gamma(z) \cos 2\varphi(z)  \\
U & \propto  \int\! dz \; n_{\rm dust} \sin^2\gamma(z) \sin 2\varphi(z)
\label{eq:stokes_dust_angles}
\end{aligned}
\end{equation}
The difference between the expressions for synchrotron and dust polarization is due to dust emission independent of the magnetic field strength. Formally, this corresponds to $\alpha=-2$ if we adopt the framework suggested in \cite{LP12}. In this paper, we explore to what extend the statistics is affected for this choice $\alpha$.

From $Q$ and $U$ one can construct $pI$ the polarized intensity
\begin{equation}
pI = \sqrt{ Q^2 + U^2 }
\label{eq:pI}
\end{equation}
where $I$ is the the intensity of the emission and $p$ is the degree of polarization, as well as determine the angle 
\begin{equation}
\cos(2 \phi) = Q/(p I) ~, \quad 
\sin(2 \phi) = U/(p I),  
\label{eq:stokes1}
\end{equation}
that defines the direction of linear polarization. Unlike the measure given by Eq.~(\ref{eq:pI}), the measures of angle defined by Eq.~(\ref{eq:stokes1}) are normalized by polarization intensities and thus much less dependent on the nature of the polarized emission.

\subsection{Variations of direction and Stokes parameters}

We first consider the statistics of the positional angle $\phi$ given in Eq.~(\ref{eq:stokes1}). This statistics is known to reflect the properties of the underlying magnetic turbulence. For instance, the structure functions of the positional angle were employed in
\cite{Fal08, 2009ApJ...706.1504H, 2009ApJ...696..567H, Laz_Pog22} for obtaining the magnetic field strength from observations for the case of sub-Alfv\'enic turbulence. In the present paper, we consider the relationship of these statistical measures with the underlying properties of super-Alfv\'enic MHD turbulence. 

From Eq.~(\ref{eq:stokes1}) we can construct the structure function for polarization angle $\phi$: 
\begin{align}
D^\phi(\mathbf{R}) &\equiv \frac{1}{4}
\left\langle \left( \frac{Q_1}{p I_1}-\frac{Q_2}{p I_2}\right)^2\right\rangle
+
\left\langle \left( \frac{U_1}{p I_1}-\frac{U_2}{p I_2}\right)^2\right\rangle 
\nonumber \\
&= \frac{1}{2} \left\langle 1 - \cos \left(2(\phi ({\bf X}_1)-\phi({\bf X}_2)\right) \right\rangle 
\label{eq:correct_dtheta}
\end{align}
where indexes $1$ and $2$ refer to two LOS separated by the 2D vector $\mathrm{\mathbf{R}}=\mathrm{\mathbf{X}_1}-\mathrm{\mathbf{X}_2}$ on the sky and the polarized intensity $pI$ is defined by Eq. (\ref{eq:pI}). This means that $D^\phi$ is readily expressed through Stokes parameters.

Note, that our measure differs from the measure introduced for the polarization angle in \citet{2009ApJ...706.1504H} by the multiplier $2$ in the cosine argument, and a factor $1/2$ in front.
The difference stems from the nature of the polarization direction that has a period of $\pi$ rather than $2\pi$. 

For sub-Alfv\'enic turbulence, this difference is of secondary importance, since it is dealing with the small-angle approximation, i.e., for $|\phi_1-\phi_2|\ll 1$, our expression given by Eq.~(\ref{eq:correct_dtheta}), as well as that of \citet{2009ApJ...706.1504H}, transfers to the "typical version" of the structure function of the angles
\begin{equation}
D^\phi(\mathbf{R}) \approx \left\langle  (\phi_1 - \phi_2)^2 \right\rangle,
\label{eq:phi_simple}
\end{equation}
that was previously explored in \cite{Fal08}.

$D^\phi$ given by Eq.~(\ref{eq:correct_dtheta}) is a general expression well-defined from observations. This is the only correct expression to apply when angle fluctuations are large, for example, for super-Alfv\'enic turbulence. If magnetic fields are completely random, for example, in the case of $M_A\rightarrow \infty$, Eq.~(\ref{eq:correct_dtheta}) asymptotes to $1/2$ value. In the presence of the mean field and for moderate $M_A$, the residual alignment of observed magnetic field directions persists, providing the saturation value less than $1/2$. We quantify the properties of $D^\phi$ below.

\subsection{Statistics of polarization degree}

The statistics of polarization degree $p$ can also be characterized by a structure function.  The corresponding normalized measure is
\begin{equation}
    D^p({\bf R})=\frac{1}{2\sigma_p^2} \left\langle (p({\bf X}_1)-p({\bf X}_2))^2\right\rangle,
    \label{eq:pol_deg}
\end{equation}
where 
%$p({\bf X})$ is the degree of polarization at the point ${\bf X}$, ${\bf R}={\bf X}_1-{\bf X}_2$, while
$\sigma_p^2$ is the variance of the degree of polarization. 

It is easy to see that Eq.~(\ref{eq:pol_deg}) differs from Eq.~(\ref{eq:correct_dtheta}) in terms of the information that can be extracted from the polarization measurements. This makes the approaches utilizing Eq.~(\ref{eq:correct_dtheta}) and Eq.~(\ref{eq:pol_deg}) complementary. In what follows, we do not consider anisotropy of the structure functions and thus discuss $D^\phi$ and $D^p$ as functions of the points separation $R$.

\section{Expectations for polarization angle statistics in high-$\beta$ medium}
\label{sec:expect}

\subsection{Model spectrum of  super-Alfv\'enic turbulence for scales less than $l_A$}

According to the model discussed above, the super Alfv\'enic $M_A > 1$ turbulence with the energy injection scale $L_{\rm inj}$ changes its behavior at the characteristic Alfv\'en scale $l_A$ given by Eq.~(\ref{l_par}). At scales $l$ exceeding $l_A$, the magnetic field is too weak and turbulence is similar to a hydrodynamic one; at scales $l < l_A$, magnetic
field backreaction is all-important. In the latter case, the turbulence can be viewed as trans-Alfvénic turbulence with the injection velocity at scale $l_A$ equal to $V_A$. The trans-Alfv\'enic turbulence follows the
GS95 cascade described by Eq.~(\ref{v_GS})

Let us now consider a single 3D cube of size $l_A \times l_A \times l_A$ that, further in the text, we term {\it $l_A$-domain}.
Within the $l_A$-domain, the cascade is anisotropic, with eddies being anisotropic with respect to the local magnetic field. In the local system of reference, where the anisotropy is calculated with respect to the magnetic field percolating the eddy (\cite{LV99, CV00, MG01}, see also Appendix A), the anisotropy increases with the decrease in the turbulence scales $\frac{l_\| }{ l_\perp } \approx
\left( \frac{l_A}{l_\perp}\right)^{1/3} $. If pointwise 3D measurements are not available, the type of anisotropy cannot be observed. Instead, in the global laboratory system of reference, e.g., the system of reference related to the mean magnetic field, the anisotropy is scale-independent and changes in the same way for all scales with $M_A$.
The wave modes $\mathbf{k}$ are determined globally in the cube $l_A$ and have a power averaged over individual eddies with different orientations. They exhibit power anisotropy primarily due to the mean magnetic field
within a cube, i.e., at the wavenumber $k_A$ defined by $k_A l_A \sim 1$.   The critical balance in terms of the wave vectors
$\frac{k_\bot}{k_\|} = \frac{l_\| }{ l_\perp } \approx
\left( \frac{l_A}{l_\perp}\right)^{1/3} = \left(k_\perp l_A\right)^{1/3}$
on the box scale gives:
\begin{equation}
\cos \theta_k \approx \sin^{2/3}\theta_k
\end{equation}
Following \cite{2002ApJ...564..291C,LP12}, the power distribution in the spectrum can be 
described by an approximate exponential model
\begin{equation}
E_{l_A}(k,\theta_k) = E(k) e^{- \frac{ |\cos\theta_k|}
{|\sin\theta_k|^{2/3}} },
\label{eq:angle_spectrum}
\end{equation}
that is applicable to wavenumbers larger than $k_A$.

It is worth noting that at the scale $l_A$ the correlations of magnetic field directions persist. Thus, $l_A$-domains can be viewed as regions with an aligned field.

\subsection{Adding contributions from $l_A$-regions along the line of sight}

In this section, we consider a qualitative picture of positional angle (PA) statistics. The first step is to evaluate what happens when the magnetic field directions of a single region, of the size $l_A\times l_A\times l_A$, are considered. Such regions we will term $l_A$-domain further in the text.
Within the $l_A$-domain, the turbulent fluctuations of the magnetic field correspond to trans-Alfv\'enic turbulence. Therefore, for scales much smaller than $l_A$, the variation of angle measured within the $l_A$ domain is $\delta \phi_l \sim \delta B/B_{d}$, where $B_d$ is the mean field in the $l_A$ domain. The latter direction defines the axis of a domain.  As a result, the PA statistics reflect the scaling of the underlying magnetic turbulence. When the point separation approaches $l_A$, the non-linearity of the relation between $\delta \phi$ and $\delta B$ should be accounted for. Nevertheless, for most of the separations less than $l_A$, the Kolmogorov approximation of the $\delta \phi$ statistics is applicable.

In the range $[l_A, L_{\rm inj}]$, where $L_{\rm inj}$ is the turbulence injection scale, the nature of the magnetic field correlation becomes different. To understand the difference, consider the eddies on scales $l_A<l<L_{\rm inj}$, that is, on scales where the kinetic energy exceeds the magnetic energy. At such scales, the magnetic field gets entangled because the magnetic stress cannot fully control hydrodynamic motions. The resulting structure of the magnetic field at $l>l_A$ depends on the spectral distribution of the magnetic field at scales larger than $l_A$. For instance, the action of a nonlinear turbulent dynamo generating a magnetic field from the magnetic field with a coherence scale less than $l_A$ results in the rising Kazantzev spectrum $E(k)\sim k^{3/2}$ (see \cite{XL16}). The effect of such a weak large-scale magnetic field on the $l_A$-domains is marginal. In contrast, if the turbulence is driven in the volume with pre-existing large-scale magnetic fields $B_0$, its effects on the orientation of $l_A$-domains can be tangible for moderate $M_A>1$.

The presence of an external magnetic field $B_0$ affects the magnetic field within the $l_A$-domains. If the angles are calculated from the direction of the external field, they depend on the mean magnetic field $B_{0}$:
\begin{equation}
    \delta \phi_l=\arctan \frac{\delta {b}_\bot}{ \delta b_\|+B_{0}},
\end{equation}
where $\delta b_\bot$ and $\delta b_\|$ are components of the magnetic field in the $l_A$ domain, respectively, perpendicular and parallel to the external magnetic field. As a result, the projected direction of the magnetic field preserves the residual alignment in the direction of the mean magnetic field, even on scales larger than $l_A$.\footnote{At scales less than $l_A$, magnetic fields preserve their coherence because the time scale of the large eddy evolution $l/v_l$ is longer than the time scale of magnetic counteraction $l_A/V_A$. In other words, although larger eddies are more dynamically powerful, they evolve more slowly than $l_A$-scale eddies.} This effect decreases $D^\phi(\infty)$ value if the external field is coherent on the injection scale. This is the case of super-Alfv\'enic turbulence driven in a volume with a magnetic mean field and having $M_A$ slightly larger than unity. As $M_A$ increases, the coherence of the magnetic field on the injection scale vanishes.

At scales larger than $l_A$, the kinetic energy drives the non-linear magnetic dynamo. The structure of magnetic fields at scales larger than $l_A$ depends on the action of the dynamo and the initial pre-existing field. As magnetic stress cannot fully constrain hydrodynamic motions, the dispersion of the angle $\delta \phi$ eventually bringing $D^\phi$ to saturation. If $\gamma$ is not $\pi/2$, the scale over which the magnetic field directions are coherent gets smaller. Indeed, the coherence of the magnetic field directions is preserved with respect to the mean field of the eddy, and the projection of this mean field changes with $\gamma$. As a result, for $\gamma$ different from $\pi/2$, the structure functions saturate at $<l_A$. 

In observations, typically, the thickness of the turbulent volume along the Line of Sight  (LOS) ${\cal L}\gg l_A$. In this situation, the axis of $l_A$-regions are randomly oriented with respect to the line of sight. This induces variations of $\gamma$ that involve the corresponding changes of $l_{A,{\rm obs}}$. For example, if for a given region $\gamma=0$, the magnetic field $B_A$ of this region is aligned with the line of sight and $l_{A,{\rm obs}}\approx 0$. For intermediate $\gamma$, $l_{A,{\rm obs}}$ varies from region to region in the interval $[0, l_A]$. It can be easily seen that for ${\cal L}\gg l_A$ and $R<l_A$, this changes the slope of the observed correlations of the magnetic field direction. In fact, for a given $R$, there are regions with such $\gamma$ that $R>l_{A,{\rm obs}}(\gamma)$. Within the adopted approximation, the PA structure functions are saturated in these regions. Due to the contributions of such regions, at sufficiently large $R$, the resulting slope of PA structure functions gets shallower than the Kolmogorov one.  

\subsection{Changes of the spectral index}

To understand the change in slope, compare the structure functions of the case at hand with a {\it test} case in which all $l_A$ domains are aligned perpendicular to the line of sight. In both cases, the structure functions are zero at $R=0$ and reach saturation for $R\approx l_A$ as for $R>l_A$. The latter effect is a consequence of the fact that, in the adopted model, in both scenarios, for any eddy along the lines of sight ${\cal L}\gg l_A$ magnetic fluctuations are not correlated for $R>l_A$. The saturation of the PA structure functions corresponds to the same value $\approx 1/2$ (see more in Appendix C). 

The test case approximately corresponds to the PA structure function with slope $R^{5/3}$. For a realistic case, at large separations, as we discussed earlier, the number of contributions along the line of sight corresponding to $R>l_A \sin \gamma$ increases. Such domains add constant contributions to the $D^{\phi}$. As a result, $D^{\phi}$ growth slows compared to the PA structure function in the test case. Thus, the slope of $D^{\phi}$ must be shallower compared to $R^{5/3}$. 

Due to the assumption that magnetic fields at $l>l_A$ are not correlated,  our approximation makes the saturation of the observed PA structure functions unrealistically abrupt at the scale $l_A$. Nevertheless, below, we will demonstrate that our model captures important properties of $D^\phi$.

The next section quantifies our considerations about the magnetic field direction statistics. To relate our study to observables, we assume that the direction of polarization reflects the direction of the magnetic field.\footnote{For the studies of turbulence with aligned dust,  this assumes a perfect grain alignment \cite{Andersson15} and homogeneous mixing of dust and gas. For studies using synchrotron polarization, the correspondence of projected magnetic field and polarization requires that the effects of depolarization and the Faraday rotation to be negligible.}

\section{Angular structure function of the projected magnetic field in super-Alfve\'enic turbulence}

\subsection{Angular structure function in $l_A$-domain}

Obtaining $Q$ and $U$ stokes parameters according to Eq.~(\ref{eq:stokes_dust}) includes LOS integration over the depth $\mathcal{L}$ that generally exceeds Alfv\'en scale, $\mathcal{L} > l_A$.  
First, let us consider how this projection works for the angular structure-function in $l_A$-domain, i.e. a small spatial volume of $l_A \times l_A \times l_A$ with a local regular field. In the next step, we will stack such volumes along the LOS to get the final expressions for the Stokes parameters. 

Local to $l_A$ domains, there is a mean magnetic field, $\overline B_A$, on top of which turbulent motions provide fluctuations with \text{rms} value $(\frac{\delta B}{B})_A \approx 1$.  Such a volume exhibits trans-Alfv\'enic turbulence with effective $M_A=1$, making applicable the LYP22 formalism developed for studies of trans-Alfv\'enic turbulence. 

LYP22 dealt with the dust polarization statistics, but the synchrotron case follows a similar formalism, as
we summarize in Appendix~\ref {Asec:PA}. There, we obtained the expression for the multipole expansion coefficients $D_s^\phi(R,\phi_R) = \sum_n D^{\phi}_n(R) e^{in\phi_R}$ of the structure function of the polarization angle when $R < l_A$. $D^\phi_s$  describes the statistics of anisotropic turbulence. However, for the purpose of this paper, we will consider only the monopole $n=0$ term that describes the orientation-averaged
structure-function.  Under this choice, Eq.~(\ref{eq:app_tildeDRgen}) reads
\begin{equation}
\label{eq:phimultipole}
D_0^\phi(R) =   \frac{\left\langle \delta B^2\right\rangle}{\overline{B}_\perp^2} \frac{\mathcal{I}_0(R)}{\mathcal{L}}
\frac{\widehat{E}_0^{2D}(\gamma)  + \widehat{E}_2^{2D}(\gamma)}{2} ~,
\end{equation}
where the corresponding functions are defined in Appendix~\ref{sec:Appendix_Dyy}. In particular,
$\mathcal{I}_0(R)$ is a function defined in Eq.~(\ref{eq:scalingfcn}) that reflects the turbulence scaling. The coefficients $\widehat{E}_p^{2D}$ are normalized POS angular harmonic decompositions of the projected power spectrum of the magnetic field that depend on the angle $\gamma$ of 3D orientation of the magnetic field in the magnetic domain relative to the LOS.

Adapting this expression to our case, we note that the role of $\mathcal{L}$ is played by $l_A$ and the projected mean field $\overline B_\perp$ in the box of size $l_A $ is
$\overline B_{A_\perp} = \sin\gamma \overline{B}_A$, where
the angle $\gamma$ varies from one $l_A$ volume to the other. To evaluate $I_0(R)$ and the $\widehat{E}_n^{2D}(\gamma)$ we use Eqs.~(\ref{eq:I0_shortR},\ref{eq:E2Dgamma}), while adopting 
the spectral model Eq.~(\ref{eq:angle_spectrum}). Thus, we obtain the short scale, $ R < l_A$  asymptotic behaviour in the form
\begin{equation}
\label{eq:phimonopole}
D_0^\phi(R) \approx \frac{\langle (\delta B/B)_A^2 \rangle}{\sin^2\gamma } A(\gamma,m) \left( \frac{R}{\l_A} \right)^{m+1}
\end{equation}
where $A(\gamma,m)$ is defined in Appendix~\ref{sec:Appendix_Dyy} and is given numerically in Figure~\ref{fig:Agamma} for the Kolmogorov scaling $m=2/3$
and spectral anisotropy modeled by Eq.~(\ref{eq:angle_spectrum}). 
By the physical meaning of the $l_A$ scale, the turbulence on this scale is trans-Alfv\'enic, i.e $\langle (\delta B/B)_A^2 \rangle \approx 1$,
though, as we will discuss below, the presence of the global mean field somewhat decreases this value. 

The angular structure function cannot exceed the value $D^\phi=1/2$,  which
corresponds to no correlations between the angles at two points.  We can estimate the effective  correlation length $R_c(\gamma)$ of polarization angles 
by extrapolating the asymptotics in Eq.~(\ref{eq:phimonopole}) to this value, $D^\phi(R_c(\gamma))=1/2$.
$R_c(\gamma)$ can be seen to strongly depend on orientation $\gamma$ of the average
magnetic field in $l_A$ domain. In the domains where it is perpendicular to LOS, $\gamma=\pi/2, A(\pi/2) \approx 0.47$, polarization angles
are correlated across the size of the domain, $R_c(\pi/2) \approx l_A $. But in the domains with the average field
nearly parallel to LOS, angle correlations disappear at much shorter separations, $R_c(\gamma\approx 0) \ll l_A$.
We adopt the following simplified
definition of $R_c$ that reflects this behaviour
\begin{equation}
R_c(\gamma) = l_A \times (\sin\gamma)^\frac{2}{m+1} ~ , 
\label{eq:Rcgamma}
\end{equation}
and to capture the saturation property of the angle structure function, we use the ansatz 
\begin{equation}
\label{eq:phimonopole_semifull}
D_0^\phi(R) \approx D_0^\phi(l_A) 
\frac{2 A(\gamma)\left(R/R_c(\gamma)\right)^{m+1}}{1+2 A(\gamma) \left(R/R_c(\gamma)\right)^{m+1}}~.
\end{equation}

The variation of $R_c$ from domain to domain will be the main reason of super-Alfv\'enic turbulent volume to exhibit polarization angle scaling
that is shallower than the projected slope $m+1$, after multiple domains, each described by Eq.~(\ref{eq:phimonopole_semifull}), are added along the LOS.

The variance of angle differences  at the domain size ,
\begin{equation}
D_0^\phi(l_A) = \frac{1}{2} \left( 1- \xi_0^\phi(l_A)\right),
\end{equation}
is reduced from the $D_0^\phi=1/2$ limit of completely disoriented polarization
directions if there are residual correlations
due to coherence of the magnetic field at scales exceeding $l_A$ such that $\xi_0^\phi(l_A)=\langle\cos 2\Delta \phi(l_A)\rangle \ne 0 $.
Such correlations come first of all from the presence of the global mean magnetic field, and, secondly, from the alignment of large 
eddies if perturbations are statistically anisotropic.  These effects are critical for sub-Alfv\'enic turbulence where the mean field is strong,  but also remain important for trans Alfv\'enic and mildly super-Alfv\'enic 
case with $M_A=[1,2]$.  A detailed study of the turbulence properties at scales $ L > l_A$ required to determine $\xi_0^\phi(l_A)$ is outside the scope of this paper. Instead, here we model the correlating effect of the global mean magnetic field approximately as
\begin{eqnarray}
\xi_0^\phi(l_A) \approx \frac{1}{1+\langle \delta B_A^2 \rangle/B_{g\perp}^2}~, \\
D_0^\phi(l_A) \approx \frac{1}{2} \times \frac{1}{1+B_{g\perp}^2/\langle \delta B_A^2 \rangle}
\label{eq:DAphi}
\end{eqnarray}
where $B_{g\perp}=B_g \sin\chi$
is the sky component of the global mean field oriented at an angle $\chi$ to LOS.
This expression matches two important limits - the absence of the effect when there is
no $B_{g\perp}$ and the increase to 100\% correlation when the projected mean magnetic field dominates the fluctuations at the $l_A$ scale. Replacing $l_A$ by
$L_{\rm inj}$ transfers us to sub-Alfv\'enic case with $B_{g\perp}^2/\langle \delta B_{\rm inj}^2\rangle \approx M_{A}^{-2} \sin^2\chi$ and $D_0^\phi(L_{\rm inj}) \propto M_A^2/\sin^2\chi$.

Putting Eqs~(\ref{eq:phimonopole_semifull}) and (\ref{eq:DAphi}) together we obtain our model for the polarization angle structure function in a $l_A$ domain:
\begin{equation}
\label{eq:phimonopole_full}
D_0^\phi(R) \approx \frac{1}{1+\zeta^2 \sin^2\chi}
\frac{A(\gamma) \left(R/l_A\right)^{m+1}}{\sin^2\gamma+2 A(\gamma) \left(R/l_A\right)^{m+1}}~.
\end{equation}
where we introduce the parameter $\zeta \equiv B_g/\sqrt{\langle \delta B_{A}^2\rangle}$ that will be important in the dicussion of LOS summation
over $l_A$ domains in the next sections.

\subsection{Summation of local contributions with no mean field }

Observations sample magnetic field in the volume much larger than $l_A$-domain.
Eq.~(\ref{eq:phimonopole_full}) presents the statistics of an individual $l_A$-domain. The summation of the contributions of such domains along the line of sign is required to obtain the structure function of the super-Alfv\'enic turbulence. We denote the latter by $D^\phi$ to distinguish it from $D_0^\phi$ for a $l_A$-domain. 

A related study has been performed in LYP22 for sub-Alfv\'enic turbulence.  The difference is that in LYP22 the system of reference was fixed and oriented with respect to the strong global mean magnetic field that sets a global system of reference in sub-Alfv\'enic regime. The Stokes parameters were calculated in relation to this chosen system of reference. In the case of super-Alfv\'enic turbulence, individual $l_A$ domains have their intrinsic fields which directions vary
from one domain to another along the LOS.

\begin{figure}[ht]
\label{fig:DfSuperA}
\includegraphics[width=0.45\textwidth]{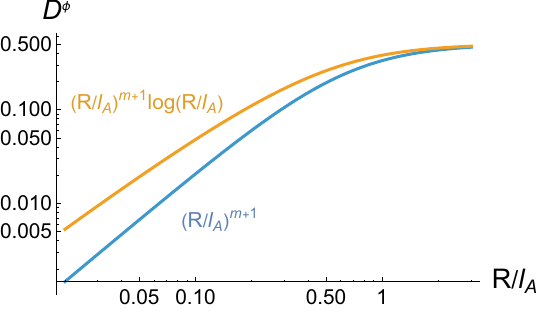}
\caption{$D^\phi$ for super-Alfv\'enic turbulence (orange line) compared to the $D^\phi$ for an individual domain (blue line). }
\end{figure}

We begin by considering the case of a magnetic field generated by a turbulent dynamo, starting with a seed field with a scale of coherence less than $l_A$. In this case, on scale $l_A$, the domains do not have any preferential contributions, and the orientation of the domains with respect to the line of sight is random. Assuming that the angles $\gamma$ of the magnetic fields of $l_A$-domains  along LOS are not correlated, and the LOS  depth is sufficiently large to contan many domains, $\mathcal{L} \gg l_A$,  the distribution of $\cos\gamma$ can be considered uniform over $[-1;1]$ range. 
{\it The resulting projected correlation function of angles is then the average correlation over all $\gamma$'s}. As an example, for the ansatz Eq.~(\ref{eq:phimonopole_full}) and $A(\gamma) \approx 1$ it is given by
\begin{equation}
\label{eq:DfSuperA}
D^\phi(R) = \left(\frac{R}{l_A}\right)^{m+1} \frac{\mathrm{ArcTanh}\left( \frac{1}{\sqrt{1+2 (R/l_A)^{m+1}}} \right)}{\sqrt{1+2 (R/l_A)^{m+1}}} ~.
\end{equation}
Fig.~\ref{fig:DfSuperA}  shows that the LOS averaging leads to a shallower slope of the structure function in comparison with the individual domain 
described by Eq.~(\ref{eq:phimonopole_full}).
At very small separations, the behavior of this structure function shows a logarithmic modification compared to the structure function of magnetic field strength, i.e.
\begin{equation}
D^\phi(R) \sim \frac{1}{2} (m+1)  \left(\frac{R}{l_A}\right)^{m+1} \ln \left(\frac{l_A}{R}\right),
\end{equation}
whereas the latter would scale $\sim R^{m+1}$. We note that the logarithmic asymptotic behaviour remains valid even when $A(\gamma)$ dependence  in Fig.~\ref{fig:Agamma} is accounted for.

\subsection{Summation in the presence of mean field}

If super-Alfv\'enic turbulence is initiated in the volume with the mean field, the average magnetic fields within domains tend to align with the global mean field. 
The degree of alignment depends on $M_A$ and vanishes for $M_A\rightarrow \infty$.  There are two reasons for the alignment; first, the global mean field
is a component of the field in the domain, and, second, the fields fluctuations  may have anisotropic properties relative to
the mean field. In the Appendix~\ref{sec:aapD} we develop a simple theory of such alignment, obtaining the distribution function $P(\theta)$ of the relative
angle between the local average and the global mean fields. There, it is shown that the distribution $P(\theta)$ depends on two parameters,  $\zeta$ 
that reflects the relative importance of the mean field,  and $\alpha$, a measure of anisotropy in the variance of perturbations at scale $l_A$.
For the spectral model Eq.~(\ref{eq:angle_spectrum}), $\alpha=0.6$ which gives a similar distribution to isotropic $\alpha=2/3$. Within the accuracy
of our discussion, we adopt $\alpha=2/3$.

The parameter  $\zeta$ has a more substantial effect for $M_A \le 2$. 
Note that for trans and sub-Alfv\'enic turbulence, where one deals with a single volume
of $L_{\rm inj}$ size, $\zeta \approx M_A^{-1}$.  In super-Alf\'enic case $\sigma_A < \sigma_{\rm inj}$, and $\zeta$ is enhanced,  
\begin{equation}
\zeta \approx \frac{\sigma_{\rm inj}}{\sigma_A} M_A^{-1}~.
\end{equation}

We do not have a first-principle theory for large scales to predict $\sigma_{\rm inj}/\sigma_A$,
but our numerical simulations  indicate in Fig.~\ref{fig:spectr_com} that super-Alfv\'enic turbulence develops nearly flat magnetic energy spectrum
$E(k) \propto k^0$ at large scales $k_{\rm inj} < k < l_A^{-1}$ and Kolmogorov $E(k) \propto k^{-5/3}$ at $k > l_A^{-1}$. Such an approximation to
spectral behaviour gives
\begin{equation}
\frac{\sigma_{\rm inj}}{\sigma_A} \approx \sqrt{\frac{5}{2}\left( 1 - \frac{k_{\rm inj}}{k_A} \right) } \approx 
\sqrt{\frac{5}{2}\left( 1 - M_A^{-3} \right)} ~,
\end{equation}
that is applicable for $M_A > 1.2$ and tends to a finite value $\sqrt{5/2}$ at large $M_A$.

\section{Expectations and numerical results}

\begin{figure*}[ht]
\label{fig:subAlf}
\centering
\includegraphics[width=1.0\linewidth]{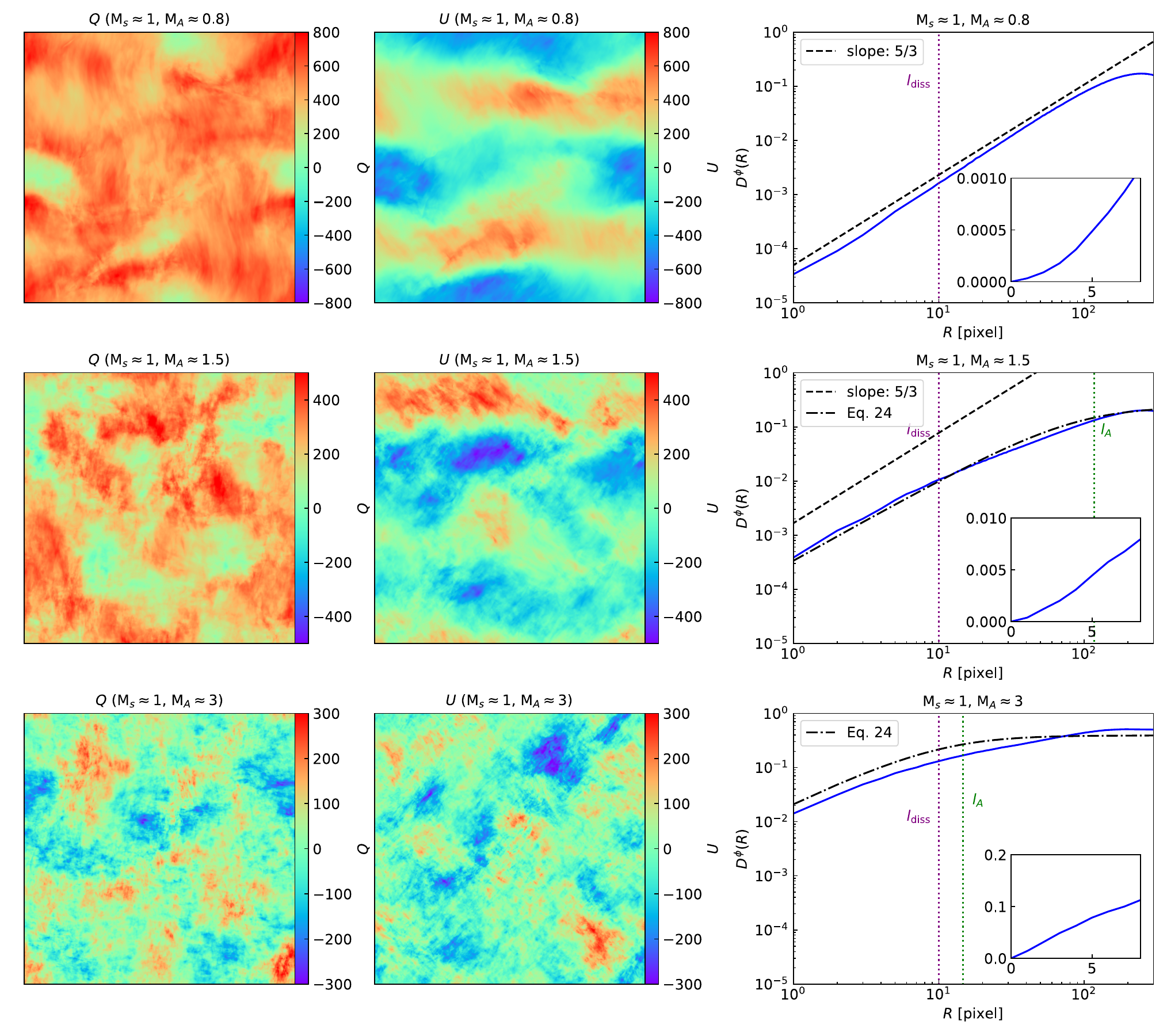}
\caption{Polarization properties when measured perpendicular to the mean magnetic field. Left and middle column: 2D maps of the Stokes parameters $Q$ and $U$ in sub-Alfv\'enic $M_A=0.8$ (top row), trans-Alfv\'enic $M_A=1.5$ (middle row), and super-Alfv\'enic $M_A=3.0$ (middle row) conditions. Right column: the structure function of polarization angle $D^\phi$. $l_{\rm diss}$ is the numerical dissipatio scale and $l_A$ is the transition scale to strong turbulence regime. The dashed line represents the expected scaling for Kolmogorov-type turbulence, and the dotted dashed line is the expectation of Eq.~\ref{eq:DfSuperA}. The subpanels illustrate the structure functions as the separation $R \to 0$.}
\end{figure*}

We plot the structure functions of the polarization directions for our numerical simulations and compare the results with our expectations (dotted lines). Note, that the our numerical simulations are performed in the box with the mean magnetic field. Thus, we have to account for its effects in Fig.~\ref{fig:subAlf} where the results obtained for $\gamma=\pi/2$ are presented. 
%The case of $\gamma=0$ is presented in Appendix \ref{appF}.
With log-log plots, it is difficult to show the behavior of structure functions at small separations. This is shown in subpanels in linear coordinates. 

We start with the case of the structure function $D^\phi$ for sub-Alfv\'enic turbulence (see \citealt{Laz_Pog22}). This allows us to better understand the differences that $M_A>1$ introduces. For $M_A=0.8$, the structure function $D^\phi_0$ is shown in Fig.~\ref{fig:subAlf}. $D^\phi$ for $M_A<1$ represents fluctuations with $\delta \phi \approx \delta B/B$. As the structure function of the magnetic field is Kolmogorov, the $D^\phi(R)$ is also following the scaling $R^{5/3}$ expected for the projected fluctuations of Kolmogorov turbulence. These results correspond to what was previously obtained for $D^\phi$ in \cite{Laz_Pog22}.

We observe that the structure functions of the polarization do not show a cut-off point at $R$ corresponding to the scale of numerical dissipation $l_{\rm diss}$. This contrasts with the energy spectrum of the observed fluctuations that shows a clear cut-off point at $k_{diss}\sim l^{-1}_{diss}$. The slope of the structure function changes from Kolmogorov $R^{5/3}$ to $R^2$ for $R<l_{\rm diss}$ (see Appendix \ref{aapC}). This slope change is small and not conspicuous in the plotted data in Fig.~\ref{fig:subAlf}.  

For $M_A=0.8$ the Stokes parameters in Fig.~\ref{fig:subAlf} demonstrate extended coherence regions for both $Q_v$ and $U_v$, representing the coherence of the magnetic field. The extent of such coherent regions decreases for super-Alfv\'enic turbulence. The visualization for $M_A=3$ demonstrates a lot of small-scale structure. These changes in polarization are also reflected in $D^\phi$.

A decorrelation of the polarization directions on the scales is present for $R>l_A$. In numerical simulations for $M_A\approx 3$, we observe small residual decorrelation for the interval $[l_A, L]$. This can be attributed to  the transport of the magnetic field by coherent large-scale hydrodynamic eddies.  Our simplified model given by Eq.~(\ref{eq:DfSuperA}) does not attempt to reproduce the slope of the functional dependence of $D^{\phi} (R)$ beyond the interval $[l_{\rm diss}, l_A]$.  

The turbulence dissipation scale $l_{\rm diss}$ and the transition scale $l_A$ are shown in Fig.~\ref{fig:subAlf}. Similarly to the sub-Alfv\'enic case, the structure functions of the magnetic field directions do not fall fast at the dissipation scale.
This complicates the estimation of the dissipation scale of turbulence using structure functions.\footnote{ Estimating the dissipation scale is possible through analysis of spectra \citep{Zhuravleva19}.}

The properties of the magnetic field direction structure functions obtained numerically correspond to our analytical predictions given by Eq.~(\ref{eq:DfSuperA}), which suggests that our simplified model captures the statistical properties of the magnetic field directions for $R<l_A$.

Our numerical simulations are performed by driving turbulence in the volume with the initial large-scale magnetic field. 
This, as we discussed earlier, modifies the statistics of magnetic fluctuations. In particular, for $M_A\approx 1.5$, $D^{\phi}$ saturates value less than $1/2$.

\section{Limitations of polarization for super-Alfv\'enic turbulence studies}

\begin{figure*}[ht]
\label{fig:synvsdust}
\centering
\includegraphics[width=1.0\linewidth]{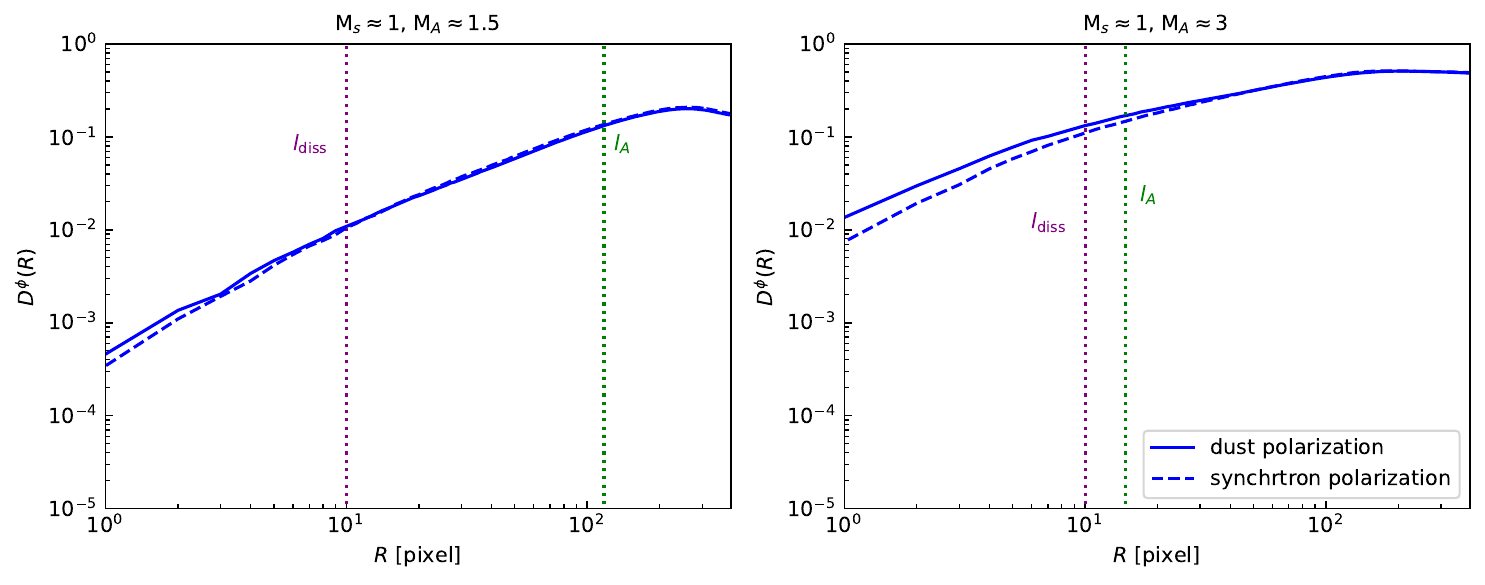}
\caption{A comparison of the polarization angle's structure function $D^\phi$ in $M_A=1.5$ (left) and $M_A=3.0$ (right) conditions. The solid line means the Stokes parameters $Q$ and $U$ are constructed in the dust polarization manner, while the dashed line represents the synchrotron polarization.}
\end{figure*}

\subsection{Relation of structure functions of dust and synchrotron polarization to projected magnetic field}

super-Alfv\'enic turbulence can be studied with synchrotron polarization, e.g., in galaxy clusters, or with dust polarization, e.g., in molecular clouds. 
Adding up polarization along the line of sight is different for synchrotron and dust emission because the synchrotron intensity is modulated by the magnetic field strength, which is absent in the case of dust. 
However, the study in \cite{LP16} analytically demonstrated a marginal dependence of magnetic field statistics on the magnetic field weighting in the emission intensity. Fig.~\ref{fig:synvsdust} shows that in both the synchrotron and dust cases, $D^\phi$ exhibits a similar behavior. Thus, $D^\phi$ that is easy to obtain from observations, presents itself as a valuable tool for synchrotron and dust polarization studies. 

The steeper slopes of the synchrotron structure function compared to the dust one are the consequence of the higher weight of fluctuations from larger scales in the fluctuations of the polarization angle. However, this does not compromise the good correspondence of the two measures.

\begin{figure}[ht]
\label{fig:directions}
\includegraphics[width=0.45\textwidth]{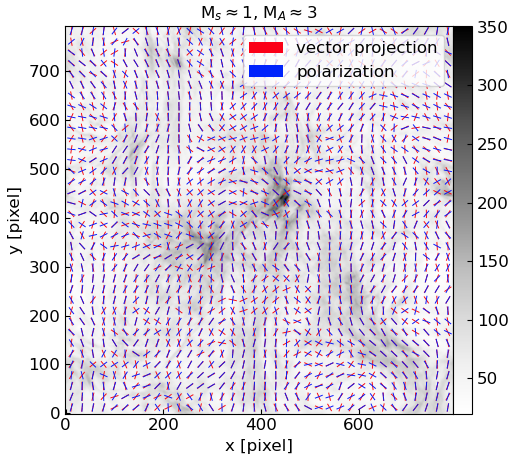}
\caption{Projected magnetic field and polarization directions for turbulence with $M_A=3$.}
\end{figure}

The next question concerns the representation of the magnetic field through polarization. The latter is widely employed as a way for tracing the Plane of Sky (POS) magnetic fields. However, the Stokes parameters add up differently from vectors along the line of sight.  We produced corresponding synthetic observations using our data cubes. Fig.~\ref{fig:directions} shows that for $M_A=3$, the structure of observed polarization and POS projected magnetic field can be quite different at the point-to-point level. This sends a warning to studies that naively identify the pattern of polarization directions with the underlying pattern of magnetic fields in astrophysical objects, e.g., in super-Alfv\'enic molecular clouds and super-Aflv\'enic media in galaxy clusters.

\begin{figure*}[ht]
\label{fig:compare}
\centering
\includegraphics[width=1.0\linewidth]{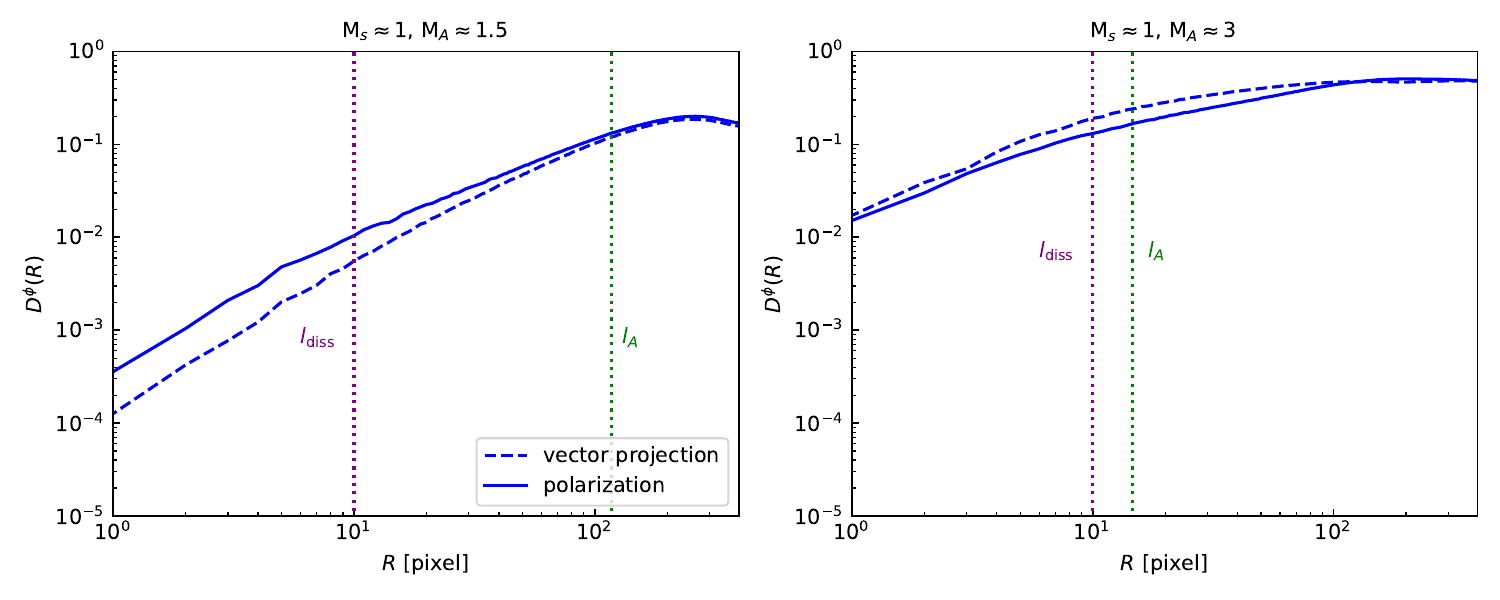}
\caption{A comparison of the polarization angle's structure functions using the Stokes parameters (solid line) and the vector projection of density-weighted magnetic fields (dashed line). }
\end{figure*}

To see to what extent these differences in direction affect the statistical properties of the maps, we plotted both the structure functions of the projected field and the structure function of polarization in Fig.~\ref{fig:compare}.  There, we observe a general correspondence of the two types of $D^{\phi}$, which demonstrates that the polarization angle structure functions can represent the statistics of the projected magnetic field despite the pointwise differences in the maps (see Fig.~\ref{fig:directions}).

\subsection{Complimentary measures}

\subsubsection{Spectrum of Directions $\cal{F_\phi}$}
\label{sec:spectrum}
\begin{figure}
\label{fig:Dspectrum}
%\centering
\includegraphics[width=0.45\textwidth]{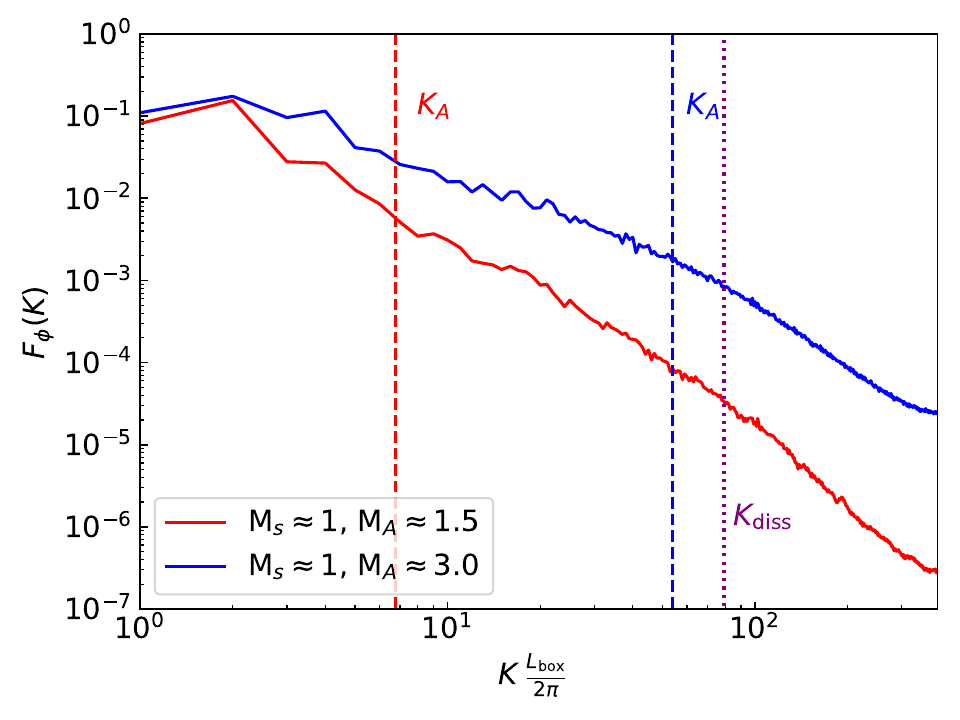}
\caption{Directional spectrum $\cal{F}_\phi$ for superAlfvenic turbulence for $M_A=1.5$ and $3$. The spectral slope at the turbulence dissipation scale is not universal. }
\end{figure}

The structure function and the spectra are complementary measures for studying turbulence. For inhomogeneous data sets, structure functions are preferable, while measured spectra can correctly represent a wider range of turbulence slopes and provide computational advantages. This motivates us to introduce a Spectrum of Directions (SD), $\cal{F}_\phi$, which is a Fourier transform of $D^\phi$. 

Using Eq.~(\ref{eq:correct_dtheta}) one can get 
that $\cal{F_\phi}$ is the sum of Fourier transforms of correlation functions $\langle \cos2\phi_1 \cos2\phi_2 \rangle$ and $\langle \sin2\phi_1 \sin 2\phi_2 \rangle$. Expressing the corresponding trigonometric functions through the Stokes parameters using Eq.~(\ref{eq:stokes1}), one gets 
\begin{equation}
    {\cal{F_\phi}}(K)=\left|{{F}}\left\{\frac{Q+iU}{\sqrt{Q^2+U^2
    }}\right\}\right|^2,
    \label{eq:Fourier}
\end{equation}
where $F$ denotes the Fourier transform and $K$ is amplitude of the Plane of Sky (POS) 2D wavevector. 

Similar to $D^\phi (R)$ given by Eq. (\ref{eq:correct_dtheta}), the directional spectrum ${\cal F_\phi}(K)$ is expressed through Stokes $U$ and $Q$, which makes both measures easy to obtain from observational data. 

Figure \ref{fig:Dspectrum} shows that the spectral slope of ${\cal F}_\phi$ changes with the value of the wavenumber corresponding to $K_{diss}\sim 1/l_{diss}$, which is in contrast to the behavior of $D^\phi$ in $R<l_{diss}$ that enters the universal $R^2$ regime (see Appendix C) when $R>4 l_{diss}$. We may observe a change in the slope of ${\cal F}_\phi$ at $K_A\sim 1/l_A$, which provides a way to determine $l_A$ from observations. However, confirming this change in slope numerically requires higher-resolution simulations.

\subsubsection{Structure function of the polarization degree}

\begin{figure*}[ht]
\label{fig:degree}
\centering
\includegraphics[width=1.0\linewidth]{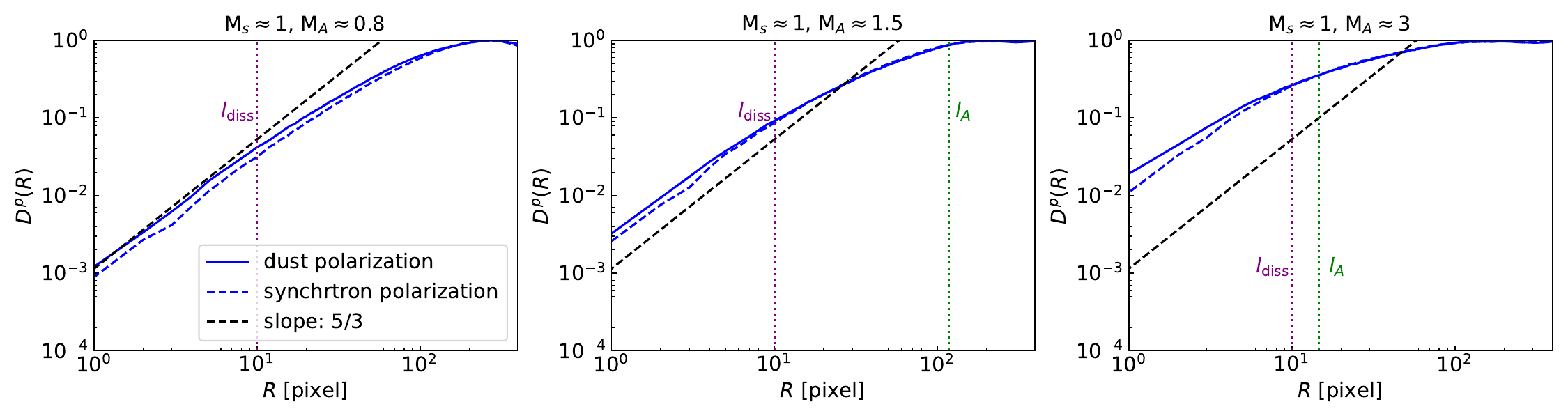}
\caption{A comparison of the polarization degree's structure function $D^p$ in $M_A=0.8$ (left), $M_A=1.5$ (middle), and $M_A=3.0$ (right) conditions. The solid line means the Stokes parameters $Q$ and $U$ are constructed in the dust polarization manner, while the blue dashed line represents the synchrotron polarization. The black dashed line represents the expected scaling for Kolmogorov-type turbulence }
\end{figure*}

Polarization varies not only in direction but also in terms of the degree of polarization. The statistics of the latter can be represented by $D^p$ given by Eq.~(\ref{eq:pol_deg}).  
The left panel of Fig.~\ref{fig:degree} illustrates the structure function of the degree of polarization $D^p$ for sub-Alfv\'enic turbulence. The calculations provided for the polarized emission of synchrotron and dust agree well with each other. 

The central and right panels of Fig.~\ref{fig:degree} demonstrate that the slope of $D^p$ becomes shallower as $M_A$ gets larger than 1. The results for dust and synchrotron polarization are very similar, especially for the $M_A=3$ case. The correspondence of $D^p$ to Kolmogorov scaling is worse compared to $D^\phi$. 
For super-Alfv\'enic turbulence, we expect that the effect of adding up $l_A$-sized magnetic domains that we described in \S \ref{sec:expect} flattens the slope of $D^p$, even though we do not have an analytical description of it. Indeed, the numerics demonstrate a flatter structure function $D^p$ $[l_{\rm diss}, l_A]$. The saturation of $D^p$ on the scale $l_A$ is more prominent compared to $D^\phi$ for $M_A=1.5$, but it is not so obvious for $M_A=3$.

\subsubsection{Synergy of measures}

We see a general correspondence between the properties of $D^p$ and $D^\phi$, which indicates that the effect of adding the contributions of the regions of $l_A$ along the line of sight that we quantified for $D^\phi$ is also present for $D^p$. An analytical study of $D^p$ properties can provide an alternative way of obtaining $l_A$ and therefore $M_A$ for super-Alfv\'enic turbulence. At present, $D^p$ can act as an auxiliary synergetic measure for studies of magnetic turbulence statistics.  This allows to use better the information available through polarization measurements.

Structure functions demonstrate a universal slope $\sim R^2$ for scales less than $4 l_{diss}$ (see Appendix C), which may not be easy to distinguish form the Kolmogorov expectations $\sim R^{5/3}$. Using the spectrum of directions ${\cal F}_\phi$ can be advantageous for determining the dissipation scale using polarization.

When the spacial variations of turbulence are of interest, combining the advantages of structure function $D^\phi$ and spectral approach using ${\cal F}_\phi$, it is possible to use a wavelet approach. However, this is beyond the scope of the present study.

\subsection{Magnetic field strength for super-Alfv\'enic turbulence: challenges and prospects}

Obtaining magnetic field strength from observations is a highly challenging astrophysical problem (see \cite{Crut10}). Davis-Chardrasekhar-Fermi (DCF) technique for measuring magnetic fields in sub-Alfv\'enic turbulence (see \cite{Davis51, ChanFer53}) as well as its modifications compares the dispersions of magnetic field directions and the velocity dispersions in molecular clouds. The original DCF is based on the equipartition of magnetic and kinetic energies at the turbulence injection scale. Later modifications (see \cite{2021A&A...647A.186S}) appeal to the $M_A^2$ disparity between the magnetic and kinetic energies. This disparity was identified in \cite{Laz_Y_V25} as arising from the velocity driving of sub-Alfv\'enic turbulence, as opposed to the driving of turbulence through magnetic fluctuations. The physical basis of the DCF is the control of fluid motions by a magnetic field. The difference in kinetic and magnetic energies by $M_A^2$ modifies, but does not disrupt this control.

The situation is radically different for super-Alfv\'enic turbulence, where the magnetic field is too weak at the injection scale to control hydrodynamic motions. Thus, the DCF approach fails for super-Alfv\'enic turbulence. In practical terms, the magnetic field measure that is employed in DCF is the magnetic field dispersion $\sim D^\phi (\infty)$. Our study in Appendix C demonstrates that $D^\phi (\infty)$ marginally depends on magnetization, i.e., on $M_A$, for $M_A>1$. Therefore, attempts to modify the DCF to measure the magnetic field strength through observations of super-Alfv\'enic turbulence are not promising.

The magnetic fields in super-Alfv\'enic turbulence are important on scales less than $l_A$.
The Differential Measure Analysis (DMA) was introduced in \cite{Laz_Pog22} to obtain detailed properties of the magnetic field through measuring $D^\phi(R)$ for $R<L$. The structure functions $D^\phi$ and the structure functions of velocity centroids:
\begin{equation}
    D^v({\bf R})=\left\langle (V({\bf X}_1)-V({\bf X}_2)^2\right\rangle,
    \label{Dv}
\end{equation}
sample magnetic fields at scales less than the injection scale, probing the magnetic field distribution. For sub-Alfv\'enic turbulence, $D^v(R)$ and $D^\phi(R)$ have the same slope over the inertial range. This opens a way to use the ratio of the aforementioned structure functions as a measure of magnetic field strength for sub-Alfv\'enic turbulence. 

One may wonder whether the DMA opens a possibility studies of magnetic field strength in super-Alfv\'enic turbulence by sampling $D^v$ and $D^\phi$ for $R<l_A$. Our present study shows that the answer to this question is negative. The addition of contributions from magnetic regions of $l_A$ size, as described in the paper, is a particular property of observational sampling of the magnetic field in super-Alfv\'enic turbulence and it differs from line of sight adding of turbulent velocities. This results in different slopes of the structure functions $D^v$ and $D^\phi$ violating the foundations of the DMA technique.

If $l_A$ is known, we know the scale at which the turbulence velocity is equal to the Alfv\'en velocity $V_A$,
i.e., $D^v=V_A^2$. Therefore, the magnetic field strength can be expressed as 
\begin{equation}
    B\approx 2 \sqrt{\pi \rho D^v(l_A)},
\end{equation}
Which is an analog of DCF formulae \cite{Davis51, CF53}, but with the velocity dispersion measured at $l_A$.

Despite the setback with the DMA, our study shows that the polarization statistics carry information about the turbulence magnetization. The revealed dependences of $D^\phi$ on $l_A$ allow to obtain $l_A$ and therefore to find $M_A$.
For instance, Eq.~(\ref{eq:DfSuperA}) relates the slope of $D^\phi$ with $l_A$. This provides a way to obtain $l_A$ from observations. Potentially, the fit observed $D^\phi$ allows one to obtain $l_A$ using a relatively small data set with $R<l_A$.  Observationally, one can obtain the injection of the turbulence scale $L$ for super-Alfv\'enic turbulence. For instance, Fig.~\ref{fig:subAlf} shows the saturation of $D^\phi (R)$ at the injection scale. Thus, using Eq.~\ref{l_A}, one can determine $M_A$, which characterizes the magnetization of the medium. $M_A$ is the key measure for the magnetization of the astrophysical medium. For example, knowledge $M_A$ is required to describe the propagation of cosmic rays \citep{YL03, LX22}. 

This way of obtaining $M_A$ is complementary to other ways of $M_A$ studies discussed in the literature (see \cite{Laz_Pog22}). Combining the value of $M_A$ with the value of the sonic Mach number $M_s$ that can also be obtained by analyzing the observational data, e.g. the statistics of synchrotron fluctuations (see \cite{2011ApJ...736...60T, 2011Natur.478..214G}), one can obtain the magnetic field strength using the approach in \citep{Laz_Y_P20, Laz_Pog22} MM2, i.e.,
\begin{equation}
    B\approx \sqrt{4\pi \rho c_s}\frac{M_s}{M_A},
    \label{mm2}
\end{equation}
where $c_s$ is the sonic speed in the medium. Note that, unlike the DCF and DMA, which are applicable only in the presence of spectroscopic data, the MM2 approach is applicable to astrophysical settings where no spectroscopic data is currently available. For instance, MM2 can be employed to galaxy cluster synchrotron polarization data.

\section{Discussion}

\subsection{Description of super-Alfv\'enic turbulence and generalization to low-$\beta$ plasmas}

The models of super-Alfv\'enic turbulence in the literature are sketchy (see \cite{BeresnyakLazarian+2019}). This study provides the first analytical expressions for quantities that are required for various branches of research. For instance, the 3D distribution of magnetic field directions over the injection scale $L$ (see Appendix \ref{sec:aapD}) is a valid quantity for the description of both plasma thermal conductivity and cosmic ray propagation in turbulent magnetic fields of galaxy clusters. For instance, in \cite{Brun_Laz07} $l_A$ was introduced as the upper limit for the effective mean free path for cosmic rays that move ballistically along magnetic field lines in super-Alfv\'enic turbulence. In the presence of partial correlation that we quantified in the paper, the effective mean free path increases making such a diffusion more efficient.

In the paper, we explain that the properties of super-Alfv\'enic turbulence can depend on the properties of the initial/seed magnetic field. In general, the turbulent magnetic dynamo in its nonlinear stage, as described in \cite{XL16}, tends to bring the magnetic and kinetic energies to equipartition, which corresponds to $M_A=1$ trans-Alfv\'enic turbulence. However, due to the relative inefficiency of the turbulent dynamo \citep{Cho_Ber09, Beres12}, which is approximately 8\% according to analytical calculations in \cite{XL16}, the observed stage of astrophysical turbulence can be far from equipartition. In this situation, the super-Alfv\'enic turbulence is influenced by the structure of the magnetic field at the moment of initialization of turbulence. For example, if for turbulence with a given $M_A$ the seed magnetic field were a random field with a correlation length less than $l_A$ corresponding to the aforementioned $M_A$, the magnetic domains remain random. In the opposite limiting case of the regular magnetic field, the domains are aligned with the initial magnetic field. This is the important addition for the understanding of super-Alfv\'enic turbulence that is introduced in our paper\footnote{If the correlation length of the initial magnetic field $l_{\rm ini, corr}$ is larger than $l_A$, then a partial correlation of domains will persist over $[l_{\rm ini,corr}, l_A]$ range. This case can be treated in analogy with our present paper, but detailed calculations are beyond the scope of this paper.} that is important for various astrophysical applications. The transient nature of super-Alfv\'enic turbulence should always be kept in mind. As turbulence evolves, $M_A$ increases due to the turbulent dynamo.

super-Alfv\'enic turbulence can be present in molecular clouds that have magnetic pressure exceeding the gaseous one, i.e., are low-$\beta$ \citep{padoan16_turb}. At the level of formation of coherent magnetic domains of the size $l_A$, the Alfv\'enic contribution is dominant and this fundamental level, the magnetic fluctuaitons of super-Alfv\'enic turbulence are the same for both low- and high-$\beta$ media \citep{CL03}. Formally speaking, the fluctuations of the magnetic field direction are determined by the Alfv\'enic cascade, and the differences found in \cite{CL03} for fast and slow modes are marginally important. However, both theoretical and numerical studies of the decomposition of MHD turbulence into fundamental modes have been centered on subsonic turbulence. In contrast, in our theoretical study, we consider high $M_A$ turbulence. For high-$\beta$ medium, this can correspond to subsonic turbulence, while for low-$\beta$ medium, the super-Alfv\'enic turbulence must be supersonic, which includes additional complications related to shocks.  

Although we avoided theoretically dealing with the study of super-Alfv\'enic turbulence in low-$\beta$ media, we provided a few general considerations that are suggestive of the applicability of our approach to low-$\beta$ media. For instance, the statistics of magnetic direction fluctuations below the $l_A$ scales are approximately Kolmogorov for both low and high $\beta$ media. In addition, in both cases, the contributions of magnetic fluctuations above the $l_A$ scale are subdominant for the structure functions of magnetic field directions. A dedicated study of the low-$\beta$ case will be done elsewhere. Nevertheless, our present study provides qualitative guidance for polarization studies of super-Alfv\'enic turbulence in molecular clouds.

\subsection{New measures for turbulence studies using polarization}

Polarization measurements have become more popular in astrophysics. The corresponding surveys encompass a wide range of wavelengths and are performed for a wide range of astrophysical media, from interstellar media and molecular clouds \cite{polar_survey21}  to relics of galaxy clusters \cite{Stuardi21}. This calls for new approaches for analysing the data. In this context, introducing new measures and studying their properties gets essential. The statistical properties of $D^\phi$ were explored for the case of sub-Alfv\'enic turbulence in \cite{Laz_Pog22}. In this paper, we extended this study for the case of super-Alfv\'enic turbulence. We showed that a non-Kolmogorov shallow-slope of $D^\phi$ indicates the super-Alfv\'enic character of turbulence there. Our study demonstrates how $M_A$ can be obtained from studying $D^\phi$.  

Studies of turbulence are also essential to understand the ecosystem of spiral galaxies \citep{2010ApJ...710..853C, Xu_Zhang16}. $D^\phi$ is a measure easy to obtain for molecular cloud polarization. The detection shallower than Kolmogorov polarization direction structure function in the presence of a Kolmogorov structure function for velocity centroids can testify that turbulence is super-Alfv\'enic. This is a good test to apply to existing data to test the in \cite{Laz_Pog22} as an alternative to the DCF, using the structure functions of magnetic field directions and velocity centroids at scales \citep{2005ApJ...631..320E, 2010ApJ...710..125E, 2014ApJ...790..130B, Vel_grad18, Huetal23II, Laz_Y_P25}.

The advantage of $D^\phi$ is that the corresponding input is readily available from observations and a search for the spectral slope that is shallower than the $5/3$ one in the range $[l_{\rm diss}, l_A]$ is straightforward. The theoretical advancement that we achieved in this paper is that we explained this effect as arising from the random orientation of the $l_A$-domains along the line of sight. Another nontrivial effect that we reported is the continuous rise of $D^\phi$ beyond the $l_A$ scale. We explain this as a consequence of either the turbulent dynamo process that generates a magnetic field at all scales smaller than $L$ or the existence of a large-scale field in the turbulent volume. The former is relevant to galaxy clusters, the latter is a feature of super-Alfv\'enic turbulence in molecular clouds. 

Our study shows the extent of the structure functions with the universal scaling $\sim R^2$ for $R<4 l_{\rm diss}$. This may make it difficult to define $l_{dis}$ with noisy observational data. The spectrum of $D^\phi$ introduced in section \ref{sec:spectrum} provides a easier way of obtaining $l_{\rm diss}$. 

The importance of studies of $D^\phi$ properties in this paper spans beyond the exploration of superAlfvenic turbulence. The measure can be used to study spectral properties of dust and synchrotron polarization. In the latter case, it is advantageous to extend the approach in \cite{LP16} to use $D^\phi$ and its Fourier transform, i.e. ${\cal F}_\phi$, to find the spectra of both magnetic field and the Faraday rotation measure. The corresponding study is provided in our next paper. 

Our in-parallel numerical study of the structure function of another polarization measure, that is, the structure function of the polarization degree $D^p$, also reveals that its slope for super-Alfv\'enic turbulence is shallower than the sub-Alfv\'enic slope $R^{m+1}$, where $m$ is the spectral slope of the underlying 3D magnetic field structure function at high wavenumbers. We can speculate that the effect of $D^p$ getting shallow has a similar origin as in the case of $D^\phi$, even though our paper does not provide an analytical proof of this statement.

The importance of our findings stems from the fact that super-Alfv\'enic turbulence is widely spread in astrophysics. With the analytical description provided, the analysis of $D^\phi$ opens a new way of exploring magnetic turbulence in galaxy clusters and other super-Alfv\'enic environments. This opens new horizons for exploring both the physical conditions and the fundamental properties of super-Alfv\'enic turbulence.

 \subsection{Getting insight into the nature of super-Alfv\'enic turbulence}

Studies of super-Alfv\'enic turbulence are challenging due to several reasons. First of all, for the same $M_A$, the structure of the magnetic field depends on whether the initial mean magnetic field was present in the simulations. If the magnetic field was amplified by a turbulent dynamo \cite{XL16} from a small-scale seed field, for the same $M_A$, the structure of the magnetic field on scales larger than $l_A$ is different from the structure of the magnetic field in turbulence driven in the presence of the large-scale field. Second, numerical simulations of super-Alfv\'enic turbulence require resolving two inertial ranges, the hydrodynamic range from $L$ to $l_A$ and the magnetically affected range from $l_A$ to $l_{\rm diss}$. Our Fig.~\ref{fig:subAlf} illustrates that this is very difficult to achieve. This makes a combination of theoretical and observational studies very advantageous. Exploring the properties of the structure functions of polarization that we performed in the paper provides a way to address this issue. 

Our present study provides an advance in understanding the nature of super-Alfv\'enic turbulence compared to the accepted views (see \cite{BeresnyakLazarian+2019}). First of all, we explored how the properties of the super-Alfv\'enic turbulence depend on the properties of the magnetic field that existed in the volume prior to turbulence driving. If the seed magnetic field is small-scale, for example, with scale less than $l_A$, a turbulent dynamo amplifies the small-scale magnetic field, creating random oriented domains of size $l_A$. If the seed field is large-scale and its energy is not negligible compared to the turbulent kinetic energy, the magnetic domains get partially aligned with the direction of the initial regular magnetic field. 
%Our study demonstrates that this alignment can be described by the Boltzman factor $\sim \exp{[-\sin^2\theta M_A^{-4}]}$. 
The model we proposed accounts for the numerical results; it can be used to explore the consequences for other processes, i.e., the propagation of heat and cosmic rays. 

\subsection{Polarization and gradients for magnetic field tracing in super-Alfv\'enic turbulence}

Polarization is an accepted way of studying the magnetic fields in sub-Alfv\'enic turbulence. This is routinely done with synchrotron polarization \cite{Beck15} and dust polarization \cite{Andersson15}. Our Fig.~\ref{fig:directions} shows that in the case of super-Alfv\'enic turbulence, the directions of observed polarization can differ significantly from those of the plane of sky projected magnetic fields. Thus, visual inspection of the polarization directions may be misleading in determining the actual magnetic field structure. However, in Fig.~\ref{fig:compare} we demonstrate that the polarization direction statistics $D^\phi$ can be similar to the projected magnetic field statistics. This means that using $D^\phi$ is meaningful in terms of understanding the statistics of the underlying magnetic field. 

We note that the directions of the magnetic field can also be obtained using the Gradient Technique (GT), which utilizes the properties of magnetized turbulence to trace magnetic fields. The corresponding measurements include the directions of velocity gradients \citep{Vel_grad18,2018MNRAS.480.1333H}, synchrotron intensity gradients \citep{synch_grad17,Hu_clusters24}, and synchrotron polarization gradients \citep{LY18polar}. 
The theory of gradients \cite{LazPog24} predicts that polarization and gradients trace magnetic fields, but the results are not identical. The difference arises from differences in adding up gradients and polarization from the fluctuations and the mean magnetic field. The gradients are not sensitive to the mean magnetic field and trace the magnetic fluctuations that are aligned with the background field. This provides for synergetic ways of using polarimetric observations and gradients. The comparison of the projected magnetic field and the direction of the gradients in \cite{Ho_super24} demonstrates that the synchrotron gradients can trace turbulence in super-Alfv\'enic turbulence better than with polarization.
We study the statistics of the gradient directions for super-Alfv\'enic turbulence in a separate paper. 

\subsection{Obtaining magnetic field strength}

When a magnetic field strength is studied using synchrotron radiation, an assumption of the equipartition of energy and energy in relativistic electrons is frequently made \cite{Beck15}. 
However, this assumption is difficult to justify, which makes the results not reliable in both sub-Alfv\'enic and super-Alfv\'enic regimes.

The assumption of equipartition of kinetic and magnetic energies is the basis of the Davis-Chandrasekhar-Fermi (DCF) approach for studying magnetic fields in molecular clouds \citep{Davis51, CF53}. For sub-Alfv\'enic turbulence this assumption is satisfied for magnetic driving of turbulence and is not satisfied for velocity turbulence driving (\cite{Laz_Y_P25}). Naturally, for super-Alfv\'enic turbulence, the kinetic energy dominates the magnetic energy. Therefore, unless $l_A$ is known, there is no way to measure magnetic field with the DCF approach that combines the dispersion of velocity that is measured through observations of Doppler-broadened lines together with the dispersion of PAs to obtain magnetic field strength cannot be justified.

The equipartition of the kinetic and magnetic energies is achieved in super-Alfv\'enic turbulence at the scale of $l_A$.
The Differential Measure Approach (DMA) was proposed as an alternative to the DCF in \cite{Laz_Pog22} for using the structure functions of magnetic field directions and structure functions of velocity centroids at a scale smaller than the injection scale. This allows for measuring magnetic field strength over regions smaller than the turbulence injection scale, i.e., to obtain magnetic field strength over localized areas. However, the DMA requires the structure functions of the velocity centroids and the structure functions of the magnetic field directions to have the same slope. This, as we see from the present paper, is not present for super-Alfvénic turbulence, making the original DMA incapable of obtaining magnetic strength either. 

Our study shows that the structure functions of magnetic field directions are sensitive to $M_A>1$. 
Finding $M_A$ opens a new way to get magnetic field strength by combining $M_A$ and $M_s$ Mach numbers, i.e., using the MM2 approach introduced in \cite{Laz_Pog22}.

\section{Summary}

In the paper above, we explore the statistics of synchrotron and dust polarization of diffuse emission arising from media with super-Alfv\'enic turbulence. We focus on the media with gas/plasma pressure exceeding magnetic pressure, i.e., the high-$\beta$ media. In other words, we do not consider highly supersonic flows with $ M_s>M_A$. Plasmas in clusters of galaxies are a prominent example of such media. Nevertheless, our results are broadly applicable to superAflfenic interstellar medium, e.g., with some limitations, to molecular clouds.\footnote{In supersonic flows, shocks are expected to modify our description of turbulence.}

Our exploration of the statistical properties of polarization was possible through the advancement of understanding of super-Alfv\'enic turbulence in the presence of the global mean field. This advance goes beyond the statistics that we are mainly dealing with in the paper, which are essential for the quantitative description of many astrophysical processes. 

We study the properties of the structure-function of the positional angle of the polarization, i.e., $D^\phi$, and compare our results with another measure of polarization, the structure function of the degree of polarization $D^p$. The advantage of $D^\phi$ is that it is readily available through the Stokes parameters of observable polarization. The measure was proven in \cite{Laz_Pog22} to be a useful tool for studies of sub-Alfv\'enic turbulence. In this paper, we explore the utility of $D^\phi$ for super-Alfv\'enic turbulence studies. We present an analytical model that explains why $D^\phi$ is shallow compared to the expectations of Kolmogorov turbulence. We compare our predictions with the results of 3D MHD numerical simulations and evaluate the prospects of studying the magnetic field properties in super-Alfv\'enic turbulence using polarization.

Our main results can be summarized as follows:
\begin{itemize}

   \item Our model of superAflvenic turbulence accounts for the magnetic field structure that consists of magnetic domains of size $l_A$. If turbulence is driven in the fluid volume with the mean field present, the domains preserve residual alignment with the mean field. We present an analytical description of the alignment and confirm our expectations with numerical simulations.

    \item For super-Alfv\'enic $M_A>1$, turbulence, the directions of polarization and the projected magnetic field may not show good pointwise correspondence, making polarization unreliable in representing of Plane-of-Sky (POS) projected magnetic field. Nevertheless, the structure functions of the polarization directions $D^\phi (R)$ and the structure functions of the POS magnetic field directions show a resonable correspondence, which justifies studying statistical properties of the magnetic field in super-Alfv\'enic turbulence using $D^\phi (R)$.
    
    \item For the underlying power-law magnetic superAflvenic turbulence, our model predicts that $D_\phi (R)$ is not a power law. The model represents the statistics of the observed polarization up to the scale of $l_A$, where the turbulent velocity of the hydrodynamic eddies becomes equal to the Alfv\'en velocity.  The residual growth of $D^\phi (R)$ for $R>l_A$ depends on the presense of the mean field in the turbulent volume.

    \item Our model predicts that the shape of $D^\phi(R)$ is a function of $M_A$ and $l_A/R$. The residual spectral slope of $D^\phi$ is shallower than the structure functions of the projected magnetic field at large lags. The analysis of $D_\phi (R)$ opens a way to determine a key scale $l_A$ at which turbulence changes its character from hydrodynamic to magnetohydrodynamic. 
    
    \item Combining the $D^\phi (R)$ and the structure function of the polarization degree $D^p(R)$ allows to utilize better the information that is available through polarization measurements.
    
    \item Both $D^\phi$ and $D^p$ functions show the universal scaling $R^{2}$ for $R<l_{\rm diss}$, which should not be confused with the slope $R^{5/3}$ arising from the projected Kolmogorov turbulence. Spectra of the functions, e.g., the spectrum of the polarization directions ${\cal F}_\phi$ that we introduced in the paper, shows an alternative way of obtaining $l_{diss}$ from observations. 
    
    \item The correspondence of the analytical predictions with the numerical simulations opens up a way of obtaining the Alfv'en Mach number $M_A$ from observations, which, in combination with the known ways of obtaining the sonic Mach number $M_s$, provides a way of recovering magnetic field strength from observations. 
\end{itemize}

{\bf Acknowledgments}
A.L. acknowledges the support of NSF grants AST 2307840. Y.H. acknowledges the support for this work provided by NASA through the NASA Hubble Fellowship grant \# HST-HF2-51557.001 awarded by the Space Telescope Science Institute, which is operated by the Association of Universities for Research in Astronomy, Incorporated, under NASA contract NAS5-26555. This work used SDSC Expanse CPU, NCSA Delta CPU, and NCSA Delta GPU through allocations PHY230032, PHY230033, PHY230091, PHY230105, PHY230178, and PHY240183 from the Advanced Cyberinfrastructure Coordination Ecosystem: Services \& Support (ACCESS) program, which is supported by National Science Foundation grants \#2138259, \#2138286, \#2138307, \#2137603, and \#2138296.  

\appendix
\counterwithin{equation}{section}

\section{Basics of turbulence within $l_a$ domain} 

In \citetalias{GS95} picture of trans-Alfv\'enic turbulence, Alfv{\'e}nic mode vectors are nearly perpendicular to the magnetic field. The corresponding motions are equivalent to eddies aligned with the magnetic field according to \citetalias{LV99}. The existence of such eddies follows from the theory of turbulent reconnection (\citetalias{LV99}, see \citealt{2020PhPl...27a2305L} for a review), which predicts that magnetic reconnection happens within one eddy turnover time. Due to this property, the magnetic field cannot constrain eddy motions in the direction perpendicular to the magnetic field. This revives the Kolmogorov picture of turbulence with the restriction that the rotation axes of the eddies are aligned with the direction of the magnetic field that surrounds the eddies. The latter is the {\it local} direction of the magnetic field. The concept of {\it local magnetic field reference} is an important addition to the original formulation of the GS95 theory, where the fluctuations are measured relative to the global mean magnetic field. This local magnetic field direction concept was proven numerically in \cite{CV00} and subsequent studies, e.g., \citep{MG01, 2002ApJ...564..291C}. The local magnetic field concept is part and parcel of the contemporary picture of the  Alfv\'enic cascade.

Due to rapid reconnection, turbulent eddies mix the magnetic field lines in a direction perpendicular to the local magnetic field. At large Reynolds numbers, the dissipation is negligible, and the energy cascades to smaller eddies whose axes are also aligned with the directions of magnetic fields in their vicinity. This induces the Kolmogorov-type condition for the flux of the cascading energy:
\begin{equation}
v_l^2/t_{{\rm casc}, l}={\rm const},
\label{casc}
\end{equation}
$v_l$ is the velocity of eddies rotating perpendicular to the magnetic field, and the scale $l_{\bot}$ and $t_{{\rm casc},l}$ are the perpendicular size of the eddy and the cascading time, respectively. The latter is the energy transfer time from an eddy of perpendicular scale $l_{\bot}$ scales to a smaller eddy and, similar to the case of Kolmogorov turbulence,
\begin{equation}
t_{{\rm casc}, l}\approx \frac{l_{\bot}}{v_l}.
\label{t_casc}
\end{equation}
The ability of free eddy rotation is ensured, as we mentioned earlier, by turbulent reconnection, which disentangles magnetic field lines.  

The parallel scale of eddies also changes in the process of cascading. Eddy rotation causes magnetic field mixing with the period $l_{\bot}/V_l$. This induces a wave with a period $l_\|/V_A$, where $V_A$ is the Alfv{\'e}n velocity. Naturally, the two periods should coincide, i.e.,
\begin{equation}
  l_{\bot}/V_l\approx  l_\|/V_A,
  \label{crit}
\end{equation}
The latter condition was termed {\it critical balance} in the \citetalias{GS95} theory of MHD turbulence.\footnote{ Originally, the critical balance was introduced for the corresponding parallel and perpendicular scales obtained relative to the {\it mean} magnetic field. In fact, the critical balance is only true in the local system of reference \cite{LV99}.} The relations between $l_{\bot}$ and $l_{\|}$ follow from the combination of Eqs. (\ref{casc},\ref{t_casc}) and (\ref{crit}), i.e. 
\begin{equation}
    l_{\|} \sim l_{\bot}^{2/3},
    \label{l_par}
\end{equation}
where $\|$ and $\bot$ scales should be calculated in terms of the {\it local direction of magnetic field} \citep{LV99, CV00,MG01}. The eddy velocities and magnetic fluctuations are Kolmogorov (see Eq.~(\ref{casc}), i.e.
\begin{equation}
    v_l\sim l_\bot^{1/3}~{\rm and}~ b_l\sim l_\bot^{1/3}
    \label{kolm}
\end{equation}
where $b_l$ is the fluctuation of the magnetic field.

In addition to Alfv{\'e}nic motions, slow and fast modes are also present in the MHD turbulence. The slow modes copy the scaling of Alfv{\'e}n modes (\citetalias{GS95},\citealt{2001ApJ...562..279L, 2002PhRvL..88x5001C,CL03}), as the Alfv\'enic cascade slaves them and imposes their structure on slow mode. This effect is present both in magnetically dominated media, i.e., in media with magnetic pressure higher than the gas pressure \cite{CL02_PRL}, and gas pressure-dominated media \cite{2001ApJ...562..279L}.
The former is usually referred to as low-$\beta$ plasma, while the latter is referred to as high-$\beta$ plasma. In the incompressible limit corresponding to $\beta \rightarrow \infty$, slow modes correspond to pure magnetic compressions. For subsonic turbulence, slow modes dominate the formation of density fluctuations and are elongated along the magnetic field with the axis ratio given by Eq.~(\ref{l_par}). The slow and fast modes play a subordinate role in our study, as they marginally affect magnetic field deviations and meandering (see \cite{LV99} which is reflected in the statistics of magnetic field direction variations that we study in this paper. Note that our description of MHD turbulence works for moderate sonic Mach number $M_s$, while the shocks play a more important role for high $M_s$.\footnote{For instance, \cite{2002PhRvL..88x5001C,CL03} demonstrated that subsonic driving provides an isotropic cascade similar to acoustic turbulence with $k^{-3/2}$ spectrum, while supersonic driving induces shock-like structures corresponding to the $k^{-2}$ spectrum \citep{Kov_Laz_2007}.}

\section{Magnetic field statistics within a $l_A$ domain}
\label{Asec:PA}

\subsection{Stokes parameters and PA structure function}
\label{Asubsec:Stokes}

LYP22 dealt with dust polarization in the volume that has a mean magnetic field and the fluctuations of the magnetic field $\delta B/B < 1$.
In the case of super-Afv\'enic turbulence, this descriptions is applicable to a $l_a^3$ individual subvolumes where the local mean
field is set by larger scales.  Let us here summarize the relevant results from LYP22.

In LP22 the $X$-axis was chosen along the sky projection of the mean magnetic field. Assuming that fluctuations of the dust density are of the same order as the fluctuations of the magnetic field $\frac{\delta n_{\rm dust}}{n_{\rm dust}} \sim \mathcal{O}\left(\frac{\delta B}{B} \right)$, in this coordinate system we find
\begin{equation}
\frac{U}{Q} \sim \frac{2}{\mathcal{L}} \int \! dz \frac{\delta B_y}{\overline{B}_x}
+ \mathcal{O} \left( \frac{\delta B^2}{B^2} \right)
\end{equation}
and
\begin{equation}
D^\phi(\mathbf{R}) \approx \frac{1}{\mathcal{L} \overline{B}_x^2}  \widetilde{D}_{yy}(\mathbf{R})
\label{eq:Dphi_def}
\end{equation}
where
\begin{equation}
 \widetilde{D}_{yy}(\mathbf{R}) \equiv 
\int \! dz \left( D_{yy}(\mathbf{R},z) - D_{yy}(0,z) \right) 
\end{equation} 
is the regularized projection of the 3D structure function $D_{yy}(\mathbf{r}) = \left\langle \left( B_y(\mathbf{r}_1) - B_y(\mathbf{r}_2) \right)^2 \right\rangle$ for the magnetic field y-component that is orthogonal to both LOS and the direction of the mean field.

\subsection{2D Projected Structure Functions}
\label{sec:Appendix_Dyy} 

Following LP12 and LYP22, the regularized projected 2D structure function  of the magnetic field components $i,j=1,2$ in MHD turbulence
can be written in Fourier space as 
\begin{align}
\widetilde{D}_{ij}(\mathbf{R}) = & \frac{1}{2\pi^2} \! \int \!\! d^2 K \left( 1 - e^{i \mathbf{K} \cdot \mathbf{R}}\right) 
\left[ \vphantom{\frac{(\mathbf{\hat K} \cdot \hat \Lambda)^2 \hat{K}_i  \hat{K}_j + \hat{\Lambda}_i \hat{\Lambda}_j - (\mathbf{\hat K} \cdot \hat \Lambda)(\hat{K}_i \hat{\Lambda}_j +   \hat{K}_j \hat{\Lambda}_i)}
{1 - \sin^2\gamma (\mathbf{\hat K} \cdot \hat \Lambda)^2} }
A(K, \sin\gamma \cos\phi_K) \!
\left( \delta_{ij} - 
\frac{\hat{K}_i \hat{K}_j + \sin^2\gamma \hat{\Lambda}_i \hat{\Lambda}_j - \sin^2\gamma \cos\phi_K (\hat{K}_i\hat{\Lambda}_j + \hat{K}_j\hat{\Lambda}_i)}
{1 - \sin^2\gamma\cos^2\phi_K}
\right)  + \right.
\nonumber \\
& \left. +  F(K,\sin\gamma\cos\phi_K)  \left( -\hat{K}_i \hat{K}_j +
\frac{\hat{K}_i \hat{K}_j + \sin^2\gamma \hat{\Lambda}_i \hat{\Lambda}_j - \sin^2\gamma \cos\phi_K (\hat{K}_i\hat{\Lambda}_j + \hat{K}_j\hat{\Lambda}_i)}
{1 - \sin^2\gamma \cos^2\phi_K}
\right) \right]
\label{eq:app_tildeDij}
\end{align}
where 2D vectors orthogonal to the LOS are introduced in capitalized notation as $\mathbf{k}=(\mathbf{K},k_z)$ and $\mathbf{r}=(\mathbf{R},z)$
while $\hat{\Lambda}$ 
 is the 2D direction of the mean field on the sky and $\cos\phi_K = \mathbf{\hat K} \cdot \hat \Lambda$.
Spectral functions $A(\mathbf{k})$ and $F(\mathbf{k})$ describe power in turbulent modes of a solenoidal field,  Alfv\'en  (A) and
orthogonal to it $F$ (see LP12).

 Considering the projected mean field to be along the $x$-direction, $\hat\Lambda=(1,0)$, the y-component structure function is 
\begin{align}
\widetilde{D}_{yy}(\mathbf{R}) = & \frac{1}{2\pi^2} \! \int \!\! d^2 K \left( 1 - e^{i \mathbf{K} \cdot \mathbf{R}}\right) 
\left[ 
A(K,\sin\gamma \cos\phi_K) \frac{ \cos^2\gamma \cos^2\phi_K}{1 - \sin^2\gamma \cos^2\phi_K}  +  \; F(K,\sin\gamma \cos\phi_K)
\frac{\sin^2\gamma \cos^2\phi_K \sin^2\phi_K }{1 - \sin^2\gamma \cos^2\phi_K} \right]
\label{eq:app_tildeDyy}
\end{align}

While LYP22 described several turbulent regimes with different distribution of power between $A$ and $F$ modes, in this paper
we consider strong turbulence regime within $l_A$-domain, which is characterized by equal power $F(\mathbf{k})=A(\mathbf{k})=E(\mathbf{k})$,
obtaining
\begin{align}
\widetilde{D}_{yy}(\mathbf{R}) = & \frac{1}{2 \pi^2} \! \int \!\! d^2 K \left( 1 - e^{i \mathbf{K} \cdot \mathbf{R}}\right) E(K,\sin\gamma \cos\phi_K) \cos^2 \phi_K ~,
\label{eq:app_tildeDijIA}
\end{align}
% One can point out that the model of strong turbulence utilizes all two degrees of freedom that 
We shall focus on the multipole coefficients $\widetilde{D}_{yy}^n (R) = 
\frac{1}{2\pi} \int_0^{2\pi} d\phi_R  \widetilde{D}_{yy}(R,\phi_R) e^{-i n \phi_R}$ for which we obtain, after performing integration
over $\phi_K$
\begin{align}
\widetilde{D}_{yy}^n (R) = & \frac{1}{2 \pi} \! \int \!\! K d K \left( \delta_{n0} - i^n J_n(K R) \right) 
\sum_{p=-\infty}^\infty E_p^{2D} (K,\sin\gamma)\; \left( \delta_{pn} + \frac{1}{2} \left(\delta_{p,n+2}+\delta_{p,n-2}\right)\right),
\label{eq:app_tildeDij0IA}
\end{align}
where $E_p^{2D} (K,\sin\gamma)$ are coefficients of 2D multipole expansion on the sky of the projected $k_z=0$
power spectrum (see \citealt{LP12,KLP17b}) with respect to angle $\phi_K$.

For astrophysical observations, only the global system of reference is available (see \citealt{LV99, CV00}), the level of anisotropy of the power spectrum is scale independent, i.e., the dependence of the power on the direction of the wavevector can be separated as
\begin{equation}
E(\mathbf{k}) = E_0(k) \widehat{E}(\widehat{\mathbf{k}} \cdot \widehat{\lambda})
, \quad \frac{1}{4\pi} \int \widehat{E}(\widehat{\mathbf{k}} \cdot \widehat{\lambda}) 
d \Omega_{\widehat{\mathbf{k}}} = 1
\end{equation}

With this factorization, 
Eq.~(\ref{eq:app_tildeDij0IA}) can be presented in the form
\begin{equation}
\widetilde{D}_{yy}^n (R) \approx  \frac{1}{2} \left\langle \delta B^2 \right\rangle
\mathcal{I}_n(R)
\sum_{p=-\infty}^\infty \widehat{E}_p^{2D} (\sin\gamma)\; \left( \delta_{pn} + \frac{1}{2} \left(\delta_{p,n+2}+\delta_{p,n-2}\right)\right)
\label{eq:app_tildeDRgen}
\end{equation}
where the formal expression for the scaling functions $\mathcal{I}_n(R)$ is
\begin{equation}
\mathcal{I}_n(R) = \frac{\pi \int \! K d K \left( \delta_{n0} - i^n J_n(K R) \right)E_0(K) }
{ \int \! k^2 dk E_0(k)}  ~~,
\label{eq:scalingfcn}
\end{equation}
in particular for the power-law spectrum $E_0(k) \propto k^{-3-m}$ in the interval $k \in (\mathcal{L}^{-1},\infty)$
\begin{equation}
\label{eq:I0_shortR}
\mathcal{I}_0(R)/\mathcal{L} \approx \frac{\pi m \Gamma\left[\frac{1-m}{2}\right]}{2^{1+m} (m+1) \Gamma\left[\frac{3+m}{2}\right]}
\left(\frac{R}{\mathcal{L}}\right)^{1+m}
\end{equation}
and $\widehat{E}_p^{2D}$ are multipole coefficients of the on-sky 2D expansion of the projected spectral angular dependence
\begin{equation}
\widehat{E}_p^{2D}(\sin\gamma) = \frac{1}{2\pi} \int_0^{2\pi} d\phi_K  e^{-i p \phi_K} \widehat{E}(\cos\phi_K\sin\gamma) 
\label{eq:E2Dgamma}
\end{equation}

Let us illustrate this result for the case of the polar angle averaged correlation, $n=0$, Kolmogorov turbulence, $m=2/3$ and the spectrum with angular dependence
given by Eq.~(\ref{eq:angle_spectrum}).  For this case $D^\phi (R)$ acquires the form
\begin{equation}
D^\phi (R) \approx  \frac{\left\langle \delta B^2 \right\rangle}{\overline{B}_\perp^2}
A(\gamma)
\left(\frac{R}{\mathcal{L}}\right)^{\frac{5}{3}}
~,\quad A(\gamma) = \left. \frac{\pi m \Gamma\left[\frac{1-m}{2}\right]}{2^{2+m} (m+1) \Gamma\left[\frac{3+m}{2}\right]} \right|_{m=2/3} \left( \widehat{E}_0^{2D} + \widehat{E}_2^{2D} \right) \approx 1.17 
\left( \widehat{E}_0^{2D} + \widehat{E}_2^{2D} \right)
\label{eq:Dphi_Agamma}
\end{equation}
where we plot $A(\gamma)$ in Fig.~\ref{fig:Agamma} below.

\begin{figure*}
\label{fig:Agamma}
\centering
\includegraphics[width=0.45\textwidth]{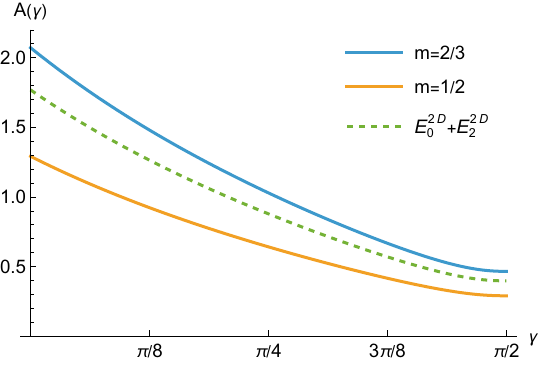}
\caption{Solid lines: Amplitude factor $A(\gamma)$ of the angle structure function in Eq.~(\ref{eq:Dphi_Agamma})
for $m=2/3$ and $m=1/2$. Dashed line: $m$ independent geometrical factor $\widehat{E}_0^{2D} + \widehat{E}_2^{2D}$ due to anisotropy of the turbulence.}
\end{figure*}

\section{Projected structure function for the dissipation range}
\label{aapC}

Let us consider how dissipation at short scales affect the scaling of the projected structure functions.  We look at a simple model where the
3D power spectrum of the turbulence is power-law subject to a Gaussian cutoff at dissipation scale $l_{\rm diss}$ 
\begin{equation}
E_0(k) = A k^{-3-m} e^{-k^2 l_{\rm diss}^2}
\end{equation}
The monopole scaling behaviour of the projected structure function is
given by $\cal{I}_0(R)$. With normalization factor omitted,  Eq.~(\ref{eq:scalingfcn}) gives
\begin{equation}
\mathcal{I}_0(R) \sim \int \! K d K \left( 1 - J_0(K R) \right) K^{-3-m} e^{-K^2 l_{\rm diss}^2} 
= \frac{\Gamma\left(\frac{1-m}{2}\right)}{1+m}\left(L_\frac{1+m}{2}\left(-\frac{l_{dis}^2}{4}\right)-1\right)
\label{eq:diss_scaling}
\end{equation}
where $L_\alpha(x)$ is a fractional Laguerre function, which is an extension of Laguerre polynomials for non-integer index. This
is not a well-known function, but what we need is its asymptotics at $ R \ll l_{\rm diss}$ and $ R \gg l_{\rm diss}$, which we can obtain
via Mathematica computer algebra software to get
\begin{eqnarray}
I_0(R) &\sim \frac{1}{4} \frac{\Gamma\left(\frac{1-m}{2}\right)}{1+m} L_\frac{m-1}{2}^1 (0)  \left(\frac{R}{l_{\rm diss}}\right)^2  & \quad R \ll l_{diss} 
\\
I_0(R) &\sim \frac{2^{-m}}{(1+m)^2} \frac{\Gamma\left(\frac{1-m}{2}\right)}{\Gamma\left(\frac{1+m}{2}\right)}  \left(\frac{R}{l_{\rm diss}}\right)^{1+m}  & \quad R \gg l_{diss}
\end{eqnarray}
where the associated fractional Laugerre function evaluated at zero $ L_\frac{m-1}{2}^1 (0)$ varies with index $m$ from $1/2$ at $m=0$ to $1$ at $m=1$.  

This demonstrates that at scales below the dissipation scale, the projected structure function acquires a universal
quadratic behavior. In contrast, on large scales it follows that law $R^{1+m}$, assuming $ m < 1$.\footnote{For $ m\ge 1$ the spectrum is too steep and leads
to saturation of the behavior of the projected structure function as $R^2$.}  What is interesting is the scale of transition from one regime to another, which can be found by equating the two asymptotics
\begin{equation}
R_{\rm trans} = \left( \frac{2^{2-m}} {(1+m) \Gamma\left(\frac{1+m}{2}\right)  L_\frac{1+m}{2}^1 (0) } \right)^{\frac{1}{1-m}} l_{\rm diss} \approx 4 l_{\rm diss}
\end{equation}
The numerical factor is practically constant, varying from $\approx 4.5$ at $m=0$ to $\approx 4.07$ at $m \to 1$.  In particular,
for Kolmogorov index $m=2/3$ we have $R_{\rm trans} \approx 4.15 l_{\rm diss}$.  Thus, in projected 2D structure functions, we expect the transition
to the dissipation regime to occur at approximately four times the Gaussian dissipation scale in the 3D spectrum. For Kolmogorov turbulence studied with noisy observational data, distinguishing the $R^2$ and $R^{5/3}$ spectra may not be easy.

\section{Angular distribution of the 3D magnetic field}
\label{sec:aapD}

In this appendix, we consider the distribution of the directions of the local magnetic field $B$ averaged in a volume of linear size $L$ in the 
presence of the global mean field $\overline B$. The knowledge of this distribution is essential for several branches of research, e.g., for cosmic ray propagation. For the problem at hand, finding this distribution is an important step for determining the correlation of projected magnetic field angles.

We shall consider the local field $B$ to be Gaussian-distributed.  In the frame where $z$ axis is aligned with $\overline{B}$,  the distribution function of $B$ field is given in polar coordinates $B_\perp = B \sin\theta$ and $B_\parallel = B \cos\theta$ as:
\begin{equation}
P(B,\theta) dB \sin\theta d\theta = \sqrt{\frac{2}{\pi}}
\frac{1} {\sigma_\parallel \sigma_\perp^2} 
\exp \left( -\frac{(B \cos\theta-\overline{B})^2}{2 \sigma^2_\parallel}-\frac{B^2 \sin ^2 \theta}{\sigma_\perp^2}\right)
B^2 dB \sin\theta d\theta
\end{equation}
where $\theta$ is and angle between $B$ and $\overline{B}$, since one expects the distribution to be uniform \textit{wrt} the azimuthal angle
around $\overline{B}$.  The variances of the fluctuations that are parallel,  $\sigma^2_\parallel$,  and perpendicular, $\frac{1}{2} \sigma_\perp^2$, to
$\overline{B}$ are potentially different, signifying the statistical anisotropy of the perturbations.

The angular distribution is obtained by performing the integration over the magnitude $B$
\begin{equation}
P(\theta)  =   \sqrt{\frac{2}{\pi \alpha^2(1-\alpha)}}
\int_0^\infty \zeta^2 d \zeta
\exp \left( -  \frac{(\zeta \cos\theta-\beta)^2}{2 (1-\alpha) } - \frac{\zeta^2 \sin ^2 \theta}{\alpha} \right)
\label{eq:Ptheta3D}
 \end{equation}
that demonstratively depends on two dimensionless parameters - the ratio of the global mean field to the \textit{rms} magnetic fluctuation at scale $L$, $\beta=\overline{B}/\sigma$, 
where $\sigma^2 = \sigma_\parallel^2 + \sigma_\perp^2=\langle |B-\overline B|^2 \rangle$, and
the relative contribution of perpendicular $B$-field components to the total variance,  $\alpha=\sigma_\perp^2/\sigma^2$.  The value
$\alpha=2/3$ corresponds to isotropic perturbations. The parameter $\alpha$ lies in the range $0 \le  \alpha \le 1$.

The general analytic result for the integral in Eq.~(\ref{eq:Ptheta3D}) exists, but is cumbersome to reproduce here. Instead, we explore two limiting cases, namely, for the anisotropic turbulence with a negligible mean magnetic field and isotropic distribution of turbulence with non-zero magnetic field. These cases allow us to evaluate the effects of turbulence anisotropy and the effect of the mean field separately.

For the case of MHD turbulence,
the anisotropy of magnetic field perturbation is intrinsically linked to the presence of the mean field that specifies the preferred direction \citep{LP12}.
However, in other contexts, e.g. in the case of turbulent dynamo (\cite{XL16}), one may have anisotropic fluctuations, $\alpha \ne 2/3$, while the mean field is negligible, $\beta \approx 0$.  In this limit
\begin{equation}
P(\theta) = \frac{2 \sqrt{2} (1-\alpha) \sqrt{\alpha }}{(2-\alpha+(3 \alpha -2) \cos 2\theta )^{3/2}} 
\end{equation}
which becomes uniform  $P(\theta)=1/2$ if $\alpha=2/3$. The case of a zero-mean magnetic field is relevant to the cosmic ray propagation in galaxy clusters
\citep[see][]{Brun_Laz07} or
when one studies the gradients of the observable quantities \citep[see][]{2020MNRAS.496.2868L,LazPog24}.

\begin{figure}
\label{fig:Ptheta3D_Bmean}
\includegraphics[width=0.45\textwidth]{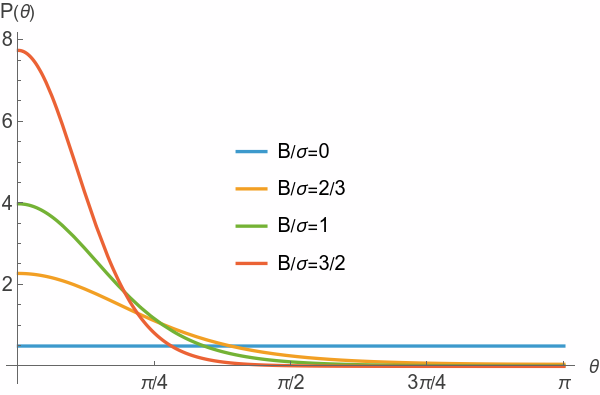}  \hfill
\includegraphics[width=0.45\textwidth]{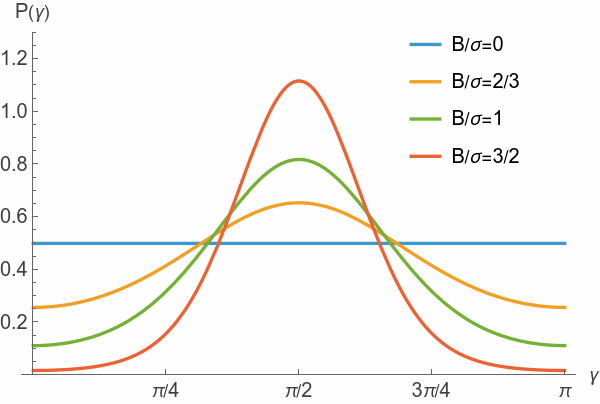}
\caption{Left: Angular distribution function $P_\theta$ 
of the 3D magnetic field orientation in the turbulent volume for different values of the mean magnetic field, for $\alpha=2/3$. 
If the mean magnetic field is oriented along the LOS, this coincides with $P_\gamma$, distribution with the respect to the coordinate angle $\gamma$. Right: Distribution
with respect to the coordinate angle $\gamma$ of magnetic in the turbulent volumes, averaged over polar orientation $\phi$,  when the mean field is perpendicular to LOS.}
\end{figure}

On the other hand, in the approximation where the anisotropy of the fluctuations is neglected,  
 $\alpha=2/3$, and the following expression
\begin{equation}
P(\theta) = \sqrt{\frac{3}{2\pi}}e^{-\frac{3 \zeta ^2}{2}} \zeta  \cos \theta+\frac{1}{2} e^{-\frac{3}{2} \zeta ^2 \sin^2\theta }
   \left(1+3\zeta ^2 \cos ^2\theta\right)
   \left(1+\mathrm{erf}\left(\sqrt{\frac{3}{2}}\zeta  \cos \theta \right)\right) ~,
\end{equation}
describes the non-uniform distribution of the directions that comes just from the mean field. Figure~\ref{fig:Ptheta3D_Bmean} shows the sample distribution for the range of $\zeta$. This distribution can characterize the magnetic field directions in the magnetized astrophysical environments. 

%\clearpaged 
\bibliographystyle{aasjournal}

\bibliography{refs.bib}
\label{lastpage}

% No idea why it goes wrong.
\end{document}